\documentclass[preprint,authoryear,5p,times,twocolumn]{elsarticle}

\usepackage{graphics}
\usepackage{color}
\usepackage{float,afterpage,longtable,amssymb,amsmath,eucal}
\usepackage[pdfborder={0 0 0}]{hyperref}
\definecolor{WDcol}{rgb}{0.0,0.0,0.0}
\definecolor{WDcolb}{rgb}{0.0,0.0,0.0}
\newcommand\rev[1]{{\color{WDcol}#1}}
\newcommand\revb[1]{{\color{WDcolb}#1}}

\usepackage{journals}

\newcommand{\vna}{\vec{\nabla}}
\newcommand{\lp}{\left(}
\newcommand{\rp}{\right)}

\journal{Icarus}
\title{\revb{Reversal} and amplification of zonal flows by boundary enforced thermal wind}
\author[leeds,mpi]{W.~Dietrich\corref{cor1}}
\ead{w.dietrich@leeds.ac.uk}
\cortext[cor1]{Corresponding author, w.dietrich@leeds.ac.uk}
\address[leeds]{Department of Applied Mathematics, University of Leeds, Leeds LS2 9JT, United Kingdom}
\author[mpi]{T.~Gastine}
\author[mpi]{J.~Wicht}
\address[mpi]{Max Planck Institute for Solar System Research,
Justus-von-Liebig-Weg 3, 37077 G\"ottingen, Germany}

\def\vec#1{\ensuremath{\mathchoice{\mbox{\boldmath$\displaystyle#1$}}
{\mbox{\boldmath$\textstyle#1$}}
{\mbox{\boldmath$\scriptstyle#1$}}
{\mbox{\boldmath$\scriptscriptstyle#1$}}}}

\newcommand{\refp}[1]{(\ref{#1})}

\newcommand{\uvec}{\mbox{$\vec{u}$}}
\newcommand{\uvecp}{\mbox{$\vec{u^\prime}$}}

\newcommand{\be}{\begin{equation}}
\newcommand{\ee}{\end{equation}}
\newcommand{\bea}{\begin{eqnarray}}
\newcommand{\eea}{\end{eqnarray}}
\newcommand{\bel}[1]{\begin{equation}\label{#1}}

\newcommand{\uvr}{\hat{\mathbf r}}

\newcommand{\E}{\mbox{E}}

\newcommand{\F}{\mathbf{F}}

\newcommand{\corr}{C_{s\phi}}

\newcommand{\geos}[1]{\left\langle #1 \right\rangle}
\newcommand{\ageos}[1]{\left\{ #1 \right\}}

\newcommand{\qs}{q^\star}

\newcommand{\Figref}[2][]{Fig.~\ref{#2}#1}
\newcommand{\figref}[2][]{fig.~\ref{#2}#1}

\newcommand{\eqnref}[2][]{eqn.~\ref{#2}#1}

\newcommand{\secref}[1]{section \ref{#1}}

\newcommand{\tabrefp}[2][]{(#1 table \ref{#2})}

\newcommand{\Tabref}[1]{Table {\ref{#1}}}

\begin{document}
\begin{abstract} 

Zonal flows in rapidly-rotating celestial objects such as the Sun, \revb{g}as or \revb{i}ce \revb{g}iants form in a variety of surface
patterns and amplitudes. Whereas the differential rotation on the Sun, Jupiter \revb{and} Saturn features a super-rotating 
equatorial region\revb{,}  the \revb{ice giants,} Neptune and Uranus \revb{harbour} a\revb{n equatorial jet slower than the planetary rotation.} Global numerical models 
covering the optically thick, deep-reaching and rapidly rotating convective envelopes \revb{of gas giants reproduce} successfully the prograde jet at the equator.   
In such models, convective columns shaped by the dominant Coriolis force typically exhibit a consistent prograde tilt. Hence 
angular momentum \revb{is pumped} away from the rotation axis via Reynolds stresses.
Th\rev{o}se models are found to be strongly geostrophic, hence a modulation of the zonal flow
structure along the axis of rotation, e.g. introduced by persistent latitudinal temperature gradients, \revb{seems of minor importance}. Within \rev{our} study we stimulate these thermal gradients and the resulting ageostrophic flows by applying an axisymmetric and 
equatorially symmetric outer boundary heat flux anomaly ($Y_{20}$) with variable amplitude and sign. Such a forcing pattern
mimics the thermal effect of intense solar or stellar irradiation. Our \rev{results} suggest that \revb{the} ageostrophic flows are linearly amplified with 
the forcing amplitude $\qs$ \revb{leading to a more pronounced dimple of the equatorial jet (alike Jupiter)}. The geostrophic flow contributions, however, are suppressed for weak $\qs$, but inverted and 
re-amplified once $\qs$ exceeds a critical value. The inverse \revb{geostrophic} differential rotation is consistently maintained by now also 
inversely tilted columns and reminiscent of zonal flow profiles observed for the \revb{i}ce \revb{g}iants. Analysis of the main force balance 
and parameter studies further foster these results.
\end{abstract}

\begin{keyword}
Zonal Flows \sep Geostrophy \sep Thermal winds \sep Heat Flux Anomalies
\end{keyword}

 \maketitle
%\linenumbers
\section{Introduction}

Zonal flows are
an essential part of the dynamics in the gaseous or
liquid envelopes of rotating celestial objects such as the
sun or giant planets.
Contrary to the smaller scale non-axisymmetric
\revb{flows}, zonal \revb{flows} are very persistent over time.
For the gas planets in our solar system \revb{surface zonal flows} have been inferred by tracking
cloud features \citep[e.g.][]{Sanchez00,Porco03,Porco05,Vasavada05}.
On Jupiter and Saturn a \revb{strong} prograde (or eastward) equatorial
jet is flanked by several alternating secondary jets at higher latitudes.
\revb{Additionally, Jupiter's main equatorial wind belt shows a pronounced dimple, where the jet amplitude at the equator is $30\%$ weaker than the surrounding main maxima \citep{Gastine2013}.}
\revb{The s}urface zonal wind profiles of Uranus and Neptune are very different, a broad
retrograde equatorial jet \revb{and} two large prograde bands at \revb{mid to higher}
latitudes \citep[e.g.][]{Sromovsky1993,Hammel2005}.
Zonal wind \revb{speeds are typically} characterised \revb{relative to the planetary rotation expressed as} the Rossby number $Ro = u_\phi
/ r_p \Omega$, where $u_\phi$ is the azimuthal velocity, $r_p$ is the planetary
radius and $\Omega$ is the \revb{angular frequency}. The peak
equatorial \revb{velocities observed} for Jupiter, Saturn, Uranus and Neptune
are then $Ro_e=0.012, 0.045, -0.08, -0.15$, respectively \citep{Aurnou2007}.

% 1) Shallow
Two competing types of models try to explain these observations. In the
'shallow' models zonal flows are driven by turbulence in a quasi-twodimensional layer.
\revb{In general,} shallow models neglect the deeper dynamics, which is somewhat hard to
motivate for \revb{the massive atmospheres of} giant planets. \revb{They typically} include \revb{crucial} physical processes like radiative transfer more
relevant for the very outer \revb{regions of} the atmosphere.
Earlier models in this category managed to reproduce multiple jet systems
with a dominant equatorial jet \citep[e.g.][]{Williams1978,Cho1996}.
However, only the more recent approaches show \revb{the correct prograde direction} of the equatorial jet
as observed on Jupiter and Saturn. E.g.~\revb{\citet{Lian2010} and \citet{Liu2011} extended the models with additional heat sources originating from e.g. 
condensation of water \revb{vapour} or solar irradiation}.

% 2) Deep rot dom.
Jupiter and Saturn-like zonal wind systems can also be naturally maintained
by deep-seated 3D \revb{convective} motions in \revb{rapidly} rotating
spherical shells.
Under the influence of a dominant Coriolis force, convection takes the form of large-scale
columnar structures that are largely invariant along the axis of rotation.
The main force balance is between the pressure gradient and the Coriolis force,
fulfilling the Taylor-Proudman theorem and establishing the so-called
geostrophic state. A mean tilt of the columns in azimuthal direction
gives rise to a statistical correlation of non-axisymmetric flows and thus
leads to Reynolds stresses that drive the zonal wind system
\citep{Busse1983,Christensen2002b, Busse2002,Plaut2008}.
The typical prograde tilt of the spiralling convective columns establishes a positive flux of
angular momentum towards the equatorial region that maintains the dominant
prograde equatorial jet \citep{Zhang92,Christensen01}.

Reynolds stress is sometimes described as a cascade from smaller to larger 
scales that stops at the Rhines scale where convective eddies start to feel the Coriolis force \citep{Heimpel2005,Gastine2014}. 
Consequently, the Rhines scale determines the width of, for example, the banded zonal flows observed on Jupiter. 

Since the pioneering work on the rotating annulus by
\citet{Busse1976,Busse1982}, it is known that the boundary curvature directly controls the
tilt direction of the convective columns and therefore sets the direction of
zonal flows; the equatorial jet is always bound to be prograde unless
additional effects start to play a role. \revb{However, these studies only cover the fundamental instabilities described by linear theory.}

% 3) Deep buo. dom.
One way of inverting the zonal flow direction is to increase the Rayleigh number
to a point where buoyancy forces become larger than Coriolis forces
(i.e. \rev{in terms of the modified Rayleigh number} $\rev{Ra^\star}\gtrsim 1$). The turbulent convective motions loose their columnar \revb{structure}
and start to \rev{stir} the spherical shell more efficiently.
The 3D mixing then \revb{homogenises} the total angular momentum (fluid flow plus system rotation)
which leads to a retrograde equatorial jet and
two flanking prograde flows \citep{Gilman79,Glatzmaier82}.  
The transition between these two regimes occurs when buoyancy
and Coriolis forces are comparable, i.e. $\rev{Ra^\star}\approx1$,
independently of the density stratification and the thickness of the convective layer
\citep{Aurnou2007,Gastine2013,Gastine2013b}.
Surface zonal flow profiles maintained in such numerical models are reminiscent
to those observed for the ice giants \citep[e.g.][]{Soderlund2013}.

% 3) Ageostrophic winds
\revb{Whether pro- or retrograde equatorial jet, 3D models of rotating convection show nearly geostrophic zonal winds, that are invariant along the axis of rotation. Ageostrophic zonal flows are typically thermal winds driven by latitudinal
temperature gradients. In planetary atmospheres, such gradients are for example established by the more intense solar heating at the equator. For all the giant planets the absorbed solar irradiation contributes to a significant fraction to the total emitted energy flux \citep{Guillot2007}. For Jupiter, Saturn, Neptune and Uranus the ratios of total emitted flux to absorbed insolation are $1.67, 1.77, 2.6, 1.06$, respectively. Hence the intrinsic heat flux is of comparable magnitude of the solar irradiation for all planets except Uranus where it amounts only to a minor contribution.}

\revb{Observations by the Pioneer spacecraft \citep{Ingersoll1975} found that the latitudinal variation of the emission is rather flat in contrast to the strong horizontal variation of the solar irradiation \citep{Soderlund2013}. As an example, for Jupiter the emission profile varies by maximal $10\%$ \citep{Pirraglia1984}, whereas the solar irradiation is at least ten times higher at the equator than at the polar region \citep{vanHemelrijck1982}. }

\revb{\citet{Aurnou2008} suggests that the intrinsic heat flux has an inverse profile that equilibrates the insolation pattern.
The solar incident flux is partially reflected and partially absorbed in the outermost atmosphere depending on the albedo. When assuming that the absorbed flux is locally re-emitted without any latitudinal redistribution the inverse insolation profile directly provides the equilibrating internal flux pattern. This heat flux anomaly is roughly shaped as an axisymmetric spherical harmonic of degree two for Jupiter, Saturn and Neptune. }

Another example is \revb{the difference in convective efficiency} between the regions inside and
outside the tangent cylinder (TC) \citep{Sreenivasan2006}. The TC is an
imaginary \revb{cylindrical} surface, that touches the inner boundary at the equator.
For small to moderate Rayleigh numbers convection remains less efficient inside the TC where 
gravity acts mostly along the rotation axis. This region is therefore cooled less 
efficiently and remains hotter, establishing a latitudinal gradient in temperature.

Latitudinal temperature (or entropy) differences translate into gradients
of the zonal winds along the axis of rotation via the thermal wind balance.
\revb{This has been studied in numerous other systems.} E.g.~an imposed latitudinal entropy contrast has for example been used by
\citet{Miesch2006} to model the non-geostrophic \revb{part of the} solar differential rotation. By
imposing a small entropy contrast at the lower boundary (a flux anomaly shaped
as a \rev{zonal} spherical harmonic degree two), they manage to maintain a solar-like
differential rotation profile with a significant deviation from geostrophy
that seems compatible with \revb{helioseismology} observations \citep{Schou2002}.
The potential impact of heat flux anomalies on convection and dynamo action has
also been investigated in several dynamo models geared to model liquid
iron cores of terrestrial \revb{solar system} \citep[e.g.][]{Amit2011,Dietrich2013} \revb{and exoplanets \citep{Dietrich2016}}.
In the context of the ancient Martian dynamo, an equatorial asymmetric
heat flux anomaly is used with increased values in the southern but decreased values
in the northern hemisphere.
For example \citet{Dietrich2015} report that larger values of the
anomaly amplitude leads to fierce non-geostrophic equatorial
antisymmetric zonal flows and a hemispherical concentration of magnetic flux.
The study of \citet{Aurnou2011} explored whether it is possible
to drive flows or even a dynamo solely by boundary forcing\rev{, i.e. outer boundary heat flux anomalies
shaped as low order \revb{sectoral} or axisymmetric spherical harmonics.}
In the absence of the radial temperature gradient that could drive columnar convection,
the zonal flow is then entirely controlled by the thermal wind.

In the present study, we \revb{analyse} how geostrophic zonal flows and
thermal winds interact \revb{and to what extent they depend on each other}. The classical Boussinesq convection model with imposed
 outer boundary heat flux pattern used here allows to control the relative strength
of geostrophic and ageostrophic zonal flows.
We \revb{focus on an axisymmetric $Y_{20}$ pattern, i.e. the mean heat flux at the outer boundary is perturbed by an equatorial symmetric flux anomaly of variable amplitude. For the bulk of the models explored, the heat flux at the equator is then smaller than at the poles. An inverse pattern is also explored for reference.}
We conduct a systematic parameter survey to study the effect
of rotation rate (Ekman number), \revb{vigour} of convection (Rayleigh number)
and relative amplitude of the heat flux anomaly ($\qs$).

The paper is organised as follows. In section~\ref{sec:intro}, we
present the hydrodynamical setup and the numerical method.
Section~\ref{sec:geos} shows how mean flows are maintained in spherical shells.
Section~\ref{sec:resultshom} focuses on the numerical models with homogeneous
heat flux, while section~\ref{sec:resultsinhom} concentrates on the
influence of a heat flux perturbation on the mean zonal flows.
Whereas section~\ref{sec:tilt} discusses the thermal wind control on the tilt of the convective columns,
in section~\ref{sec:params} we present the results of the parameter study
before concluding the paper in section~\ref{sec:conclu}.

\section{Hydrodynamical setup}
\label{sec:intro}

\subsection{Governing equations}
We consider numerical simulations of convection in a spherical shell rotating
at a constant rotation rate $\Omega$ about the $z$-axis. Under the Boussinesq
approximation, the evolution is governed by a set of
non-dimensional equations for the conservation of mass, momentum and
thermal
energy:

\begin{eqnarray}
\nabla \cdot \vec{u} &=&0  \label{eq:cont} \ , \\
\frac{\partial \vec{u}}{\partial t} +\vec{u} \cdot \nabla \vec{u} &=&  -\nabla
\pi - 2 \vec{e}_z \times \vec{u} + \frac{Ra^\star}{r_o}
T \vec{r} + E\,\nabla^2 \vec{u} \label{eqnst} \ , \\
\frac{\partial T}{\partial t} + \vec{u} \cdot \nabla T &=& \frac{E}{Pr} \nabla^2T
+ \epsilon \label{eq:nrj} \ ,
\end{eqnarray}
where $\vec{u}$ is the fluid velocity, $T$ the super-adiabatic temperature,
and
$\nabla \pi$ the gradient of the non-hydrostatic pressure. $\vec{e}_z$ is the
unit vector along the axis of rotation and $\epsilon$ is a uniform heat source
density. As in previous studies \citep[e.g.][]{Christensen2002b, Aubert2005},
we adopt a dimensionless formulation using \rev{$\Omega^{-1}$} as the time unit,
the thickness of the spherical shell  $D=r_o-r_i$ as the reference length scale
and the mean heat flux at the outer boundary $q_0$ as the temperature scale
$\Delta T = q_0 D/ \rho c_p \kappa$. Here $\rho$ is the constant density\rev{,}
$c_p$ the constant \rev{specific} heat capacity \rev{and $\kappa$ the thermal diffusivity}. 
The system of equations
(\ref{eq:cont}-\ref{eq:nrj}) is \revb{governed}
by three control parameters, namely the modified flux-based
Rayleigh number $Ra^\star$ \citep[e.g.][]{Christensen2002b}, the Ekman number
$E$ and the Prandtl number $Pr$:

\begin{eqnarray}
Ra^\star &=& \dfrac{\alpha g_0 q_0}{\rho c_p \kappa \Omega^2}, \label{eqdefrastar}\\
E &=&\dfrac{\nu}{\Omega D^2}, \\
Pr&=&\dfrac{\nu}{\kappa},
\end{eqnarray}
where $\nu$ and $\kappa$ are the constant kinematic and thermal
diffusivities, $\alpha$ is the thermal expansivity and $g_0$ is the gravity at
the outer boundary. $Ra^\star$ can be related to the \revb{definition of the} classical flux-based
Rayleigh number $Ra=\alpha g_0 q_0 D^4 /\rho c_p \nu \kappa^2$ via $Ra^\star
=Ra\,E^2/Pr$. 

\revb{The flux based Rayleigh number $Ra^\star$ provides a measure of the ratio of buoyancy to Coriolis forces. However as defined in Eq.~\ref{eqdefrastar}, $Ra^\star$ provides a reference value at the outer boundary. Furthermore \citet{Gastine2013} argues, that a mid-depth value $Ra_m^\star$ is more relevant to classify the force balance. This can be based
on the mid-depth gradient of the conductive temperature $\tilde{T}$}:
\begin{equation}
Ra_m^\star=\frac{\alpha g(r_m)}{\Omega^2} \left| \frac{\tilde{T}}{dr} \right|_{r_m} .
\label{eqdefram}
\end{equation}
The conductive background state for a purely internal heated spherical shell is:
\begin{equation}
\kappa\rho c_p\frac{d \tilde{T}}{dr}= \frac{1}{3} \epsilon^\prime \left(\frac{r_i^3}{r^2} -r\right)
\end{equation}
where $\epsilon^\prime$ is the dimensional heat source density.
\revb{To avoid a mean drift of the internal temperature we require that the heat flux through the outer boundary is balanced by internal heat sources, so that}:
\begin{equation}
 \epsilon^\prime = \frac{3 r_o^2}{r_o^3-r_i^3} q_0 \ ,
\end{equation}
and hence
\begin{equation}
  Ra_m^\star = \frac{r_o}{r_m}\frac{r_m^3-r_i^3}{r_o^3-r_i^3} Ra^\star\;\;,
\end{equation}
where $r_m=(r_o+r_i)/2$ is the mid radius of the convective shell. For the aspect
ratio $a=r_i/r_o= 0.35$ and linear gravity ($g(r_m) = r_m/r_o$) assumed here 
this yields $Ra_m^\star\approx 0.41 Ra^\star$.\\

In the numerical models presented in this study, we assume stress-free
mechanical boundary conditions and a fixed heat flux at both boundaries.
The imposed bottom flux is set to zero to ensure that convection is exclusively
driven from the internal heat source.
The outer boundary heat flux is the sum of a mean \revb{contribution} $q_0$ and an axisymmetric
pattern described by the spherical harmonic
surface function $Y_{lm}$ of degree $l=2$ and order $m=0$:
\begin{equation}
q (\theta,r=r_o)= q_0 (1- q^\star Y_{20}) \ .
\label{eqanomaly}
\end{equation}
Here, $q^\star$ is the relative amplitude of the $Y_{20}$ variation, \revb{normalised}
in such a way that the equatorial heat flux vanishes for $q^\star=1$
while the polar flux is then three times the mean flux $q_0$. \revb{As discussed in the introduction this pattern attempts to mimic the fundamental effect of stellar irradiation in a simplified way}. 

 For completeness, we also
investigate models \revb{with} $q^\star < 0$ where the \revb{heat flux at the} equator \revb{is higher than} the poles. \revb{Generally, we limit the amplitude of $q^\star$ to values where the total heat flux never decreases below the adiabatic value. This seems preferable to guarantee that our model assumptions of small disturbances around an adiabatic state still holds.}

\subsection{Numerical methods}
The numerical simulations in this parameter study were computed with the
pseudo-spectral code MagIC \citep{Wicht2002, Christensen2007}. \revb{An updated version of the code can be found on \url{https://github.com/magic-sph/magic}}.  To solve the
system of equations (\ref{eq:cont}-\ref{eq:nrj}) in spherical coordinates
$(r,\theta,\phi)$, the \revb{incompressible velocity $\vec{u}$ \revb{is} represented by two scalar potentials, such that}
\begin{equation}
\vec{u} = \vna\times\lp\vna\times W \uvr\rp +
\vna\times Z \uvr,
\label{eq:decomposition}
\end{equation}
where $W$ and $Z$ are the poloidal and toroidal \revb{flow} potentials, respectively. $W$, $Z$, $\pi$ and
$T$ are expanded in spherical harmonic functions up to degree and order
$\ell_\text{max}$ and in Chebyshev
polynomials up to degree $N_r$ in radius. 

\revb{For the models explored here,} we use numerical truncations
ranging from ($N_r=49,\,\ell_{max}=96$) to ($N_r=97,\,\ell_\text{max}=213$). 
All cases have been \revb{time-integrated} over more than one
viscous diffusion time to ensure that a statistically steady state has been
reached. We fix the hydrodynamic Prandtl number to $Pr=1$, the
aspect ratio to $a=0.35$ and vary the Ekman number $E$, the modified
Rayleigh number $Ra^\star$ and the amplitude of the heat flux anomaly $q^\star$.

The Ekman numbers considered in this study span two decades from
$E=10^{-3}$ to $10^{-5}$, the Rayleigh number roughly three decades from
$Ra^\star=10^{-2}$ to $10$ and the heat flux anomaly amplitude ranges from
$q^\star=-1$ to $1$. \revb{In total, we have computed 92 different numerical models.}

\section{Geostrophic and ageostrophic zonal flows}
\label{sec:geos}

%In this section we derive the equations describing Reynolds stress \revb{and thermal winds} and
%establish the link to the correlation of non-axisymmetric flows \rev{and} thus the tilt
%of convective features.
\revb{An} analysis of the Navier-Stokes equation shows that the axisymmetric azimuthal or zonal flows
$\overline{u}_\phi$ can only be modified by three forces: the nonlinear inertial force or advection $\overline{F}_{NL}$,
the Coriolis force $\overline{F}_{C}$, and the viscous force $\overline{F}_{V}$. Buoyancy forces are purely radial
and the azimuthal pressure gradient has no axisymmetric contribution.
On time average, the remaining three zonal forces should balance:
\bel{NS1}
 \overline{F}_{NL} + \overline{F}_{C} + \overline{F}_{V} = 0 \;\;\rev{,}
\ee
\rev{where overbars generally denote azimuthal averages.}
The zonal Coriolis force is simply
\bel{CO}
\overline{F}_{C} = -\overline{u}_s\;\;,
\ee
where $u_s$ is the axisymmetric flow contribution perpendicular to the
rotation axis. For the incompressible flows considered here,  $\overline{F}_{C}$
has no geostrophic component since the net flow across the
cylinder must vanish: $\geos{\overline{F}}_C=0$. The geostrophic part is defined by a vertical
average:
\begin{equation}
 \geos{\overline{F}}  = \frac{1}{h(s)} \int_{z_{-}}^{z_{+}} \overline{F}\, dz \ ,
\label{eqheigth}
\end{equation}
where $h(s)=z_{+}-z_{-}$ is the height of the container. \revb{Outside TC eq.~\ref{eqheigth} defines one integral spanning the whole core with} $h(s)= 2z_{+}= 2 \sqrt{r_o^2-s^2}$. \revb{Inside the TC there are two integrals for northern and southern hemisphere, respectively.} \revb{Hence the ageostrophic contribution is defined by}:
\begin{equation}
 \ageos{\overline{F}} = \overline{F} -\geos{\overline{F}} \ .
\end{equation}

The geostrophic part of the time-averaged zonal Navier-Stokes equation
thus \revb{simply reads}
\bel{NS2}
 \geos{\overline{F}}_{NL} + \geos{\overline{F}}_{V} =0
 \;\;.
\ee

We start with considering the zonal contribution of the nonlinear advective force which
formulates the interplay of \revb{different} flow components\rev{:}
\bel{fp1}
\overline{F}_{NL} =
-\overline{u_s\partial_s\,u_\phi} -
 s^{-1} \overline{u_\phi\partial_\phi\,u_\phi} -
 \overline{u_z\partial_z\,u_\phi}  -
 s^{-1} \overline{u_s u_\phi}\;\;.
\ee
This can by simplified
using the \revb{incompressibility condition} $\nabla\cdot\uvec=0$
and \revb{then the geostrophic contribution yields}
\bel{fp3}
\geos{\overline{F}}_{NL} =
- \geos{s^{-2}\partial_s s^2 \left(\overline{u_s\,u_\phi}\right)} -
\left. h^{-1}\;
\left(\overline{u_z\,u_\phi}\right)\;\right|_{z_-}^{z_+}\;\;.
\ee
%For the stress-free boundaries considered here, 
\rev{W}e use the fact that
the radial component has to vanish at the boundaries which implies
\bel{bcs}
 u_z=-\tan(\theta)\;u_s=-\frac{s}{z}\;u_s
\ee
at $z_-$ and $z_+$.
Expression (\ref{fp3}) then becomes
\bel{fp4}
\geos{\overline{F}}_{NL} =
- \left< s^{-2}\partial_s \left( s^2 \overline{u_s\,u_\phi}\right)\right> +
\left. s\; h^{-1}\;
\left(z^{-1} \overline{u_s\,u_\phi}\right)\;\right|_{z_-}^{z_+}\;\;.
\ee

When taking into account that the boundaries of the
$z$-integral depend on $s$ this further simplifies to
\bel{fp5}
\geos{\overline{F}}_{NL} =
- h^{-1} s^{-2}\partial_s \left( h s^2 \geos{\overline{u_s\,u_\phi}}\right)\;\;,
\ee
an expression that only depends on the correlation of $u_s$ and $u_\phi$ integrated
over geostrophic cylinders.
%Note that the factor $h s^2$ under the differential guarantees that
%the Reynolds stress conserves the total angular momentum.

The axisymmetric non-linear inertial force can be separated in general into a contribution
due to the interaction of non-axisymmetric flow components and a
second contribution which describes the action of the meridional circulation:
\bel{Faa}
\overline{\F}_{NL}=\overline{\F}_{RS} + \overline{\F}_{AD} =
- \;\overline{\left({\uvec}^\prime\cdot\nabla\right)\;{\uvec}^\prime} -
  \left(\overline{\uvec}\cdot\nabla\right)\;\overline{\uvec}\;\;.
\ee
The former contribution is \rev{the force due to} Reynolds stress\rev{, also 
described as Reynolds stress convergence} while the latter has
been dubbed advective force by \citet{Wicht2010}.
According to \eqnref{fp5}
the azimuthal geostrophic components of these forces are
\bea
\geos{\overline{F}}_{RS}\label{spgeo}
&=&
- h^{-1} s^{-2}\partial_s \left( h s^2 \geos{\overline{u^\prime_s\,u^\prime_\phi}}\right) \\
\geos{\overline{F}}_{AD} &=&
- h^{-1} s^{-2}\partial_s \left( h s^2 \geos{\overline{u}_s\,\overline{u}_\phi}\right) \;\;.
\eea

We next turn to the viscous force, wh\revb{ose} axisymmetric azimuthal component is given by
\bel{fv1}
\overline{F}_{V} =
\E\;s^{-2} \partial_s \left( s^{3} \partial_s s^{-1} \overline{u}_\phi \right)  +
     \E\;\partial_z^2 \overline{u}_\phi\;\;
\ee
with the geostrophic contribution
\be
\geos{\overline{F}}_{V} =
\E\;\geos{ s^{-2}\partial_s \left( s^{3} \partial_s s^{-1} \overline{u}_\phi \right)}  +
    \E\;\left. h^{-1}\;
    \left(\partial_z\overline{u}_\phi\right)\right|_{z_-}^{z_+}\;\;.
\ee
When taking the $s$-dependence of the boundaries into account and
using the stress-free boundary condition $(\partial_r r^{-1}\overline{u}_\phi=0)$
this simplifies to
\bel{fv2}
\geos{\overline{F}}_{V} =\E\;
h^{-1}s^{-2}\partial_s \left( h s^{3}
\geos{\partial_s s^{-1}\overline{u}_\phi }\right) \;\;.
\ee

A comparison of \eqnref{fp5} and \eqnref{fv2} reveals that 
\bel{UT}
 \geos{\overline{u_s\,u_\phi}} =
  \geos{\overline{u_s^\prime\,u_\phi^\prime}}
+  \geos{\overline{u}_s\,\overline{u}_\phi} =
 \E\;s\;\geos{\partial_s s^{-1} \overline{u}_\phi }\;\;
\ee
is equivalent to a balance between nonlinear inertial and viscous \revb{stresses}
on geostrophic cylinders and should thus hold on time average.

An \revb{analysis} of the different zonal force cont\rev{r}ibutions in our simulations revealed
that the advective force \rev{$\geos{\overline{F}}_{AD}$} is at least one order of magnitude smaller
than the Reynolds stress \rev{convergence} \rev{$\geos{\overline{F}}_{RS}$} for all the cases explored here.
The reason is the generally weak
meridional circulation component $\overline{u}_s$ the advective force
relies on. 
\Tabref{tabflow} \revb{demonstrates that for the volumetrically and time averaged flow correlations $\cal{RS}$ and $\cal{AD}$ defined by
\begin{align}
\mathcal{RS} = \int_{r_i}^{r_o} \int_0^\pi \overline{u_s^\prime u_\phi^\prime}\,  r^2 \sin \theta \, dr \, d\theta \label{eqdefRS} \\
\mathcal{AD} = \int_{r_i}^{r_o} \int_0^\pi \overline{u}_s \overline{u}_\phi \, r^2 \sin \theta \, dr \, d\theta 
\end{align}
the axisymmetric part is indeed much smaller than the respective non-axisymmetric contributions}.
We can thus \rev{safely} neglect \rev{$\geos{\overline{F}}_{AD}$} and also the
$\geos{\overline{u}_s\,\overline{u}_\phi}$ contribution in balance $\refp{UT}$
in our attempt to understand the reason for the zonal flow inversion. 

\revb{A vertical variation of the axisymmetric, azimuthal flows arises from e.g.~from consistent temperature variations along latitude.} The $z$-variation of zonal flows $\ageos{\overline{u}_\phi}$ is \revb{dominated by} thermal
winds \revb{which are driven by consistent temperature variations along latitude}. \revb{This} can be understood by taking the azimuthal component of the curl
of the Navier-Stokes equation (Eq.~\ref{eqnst}):
\begin{equation}
 \frac{\partial \omega_\phi}{\partial t} +  \left[ \nabla\times(\vec{\omega}\times
\vec{u})\right]_\phi  = \frac{\partial u_\phi}{\partial z} - \frac{Ra^\star}{2 r_o}
\frac{\partial \rev{T}}{\partial \theta} + E\Delta \omega_\phi,
\end{equation}
where $\omega_\phi$ is the azimuthal component of vorticity $\vec{\omega} =
\vec{\nabla}\times \vec{u}$. \revb{For the geostrophic flow contributions the consistent tilt of the convective columns provides a mean azimuthal correlation and hence a sizeable Reynolds stress. Since there is no respective mechanism for the ageostrophic flow, the zonal average of the respective nonlinear advection is generally small. Since viscous effects are also small at the Ekman numbers considered here, the time-persistent mean temperature structure yields the simplified thermal balance between Coriolis and buoyancy forces: }
\begin{equation}
\frac{\partial \ageos{\overline{u}_\phi}}{\partial z} \simeq \frac{Ra^\star}{2 r_o}
\frac{\partial \overline{T}}{\partial \theta} \ .
\label{eqtw}
\end{equation}
%According to eq.~\ref{eqtw} hotter regions are linked to \revb{slower than average} 
%zonal flows, whereas cooler regions show an enhanced
%prograde zonal flow.

%\revb{To summarise,} the equations demonstrate \revb{that} the driving of geostrophic zonal flows
%via Reynolds stresses requires a statistical correlation
%between the non-axisymmetric flow components $u_s^\prime$ and
%$u_\phi^\prime$. This correlation is typically provided by a
%systematic tilt of convective columns.
%Persistent ageostrophic axisymmetric zonal flows, on the other hand,
%are mostly driven via persistent latitudinal temperature gradients.
%Here we enforce these gradients via the imposed outer boundary
%heat flux pattern. \revb{This suggests, that geostrophic and ageostrophic
%zonal flows are independent of each other. Within our numerical experiments
%we stimulate the latitudinal temperature gradient via the imposed outer boundary
%heat flux pattern and show that there is an unexpected, but clearly identifiable relation amongst them.} 

\section{Results for homogeneous outer boundary heat flux}
\label{sec:resultshom}
\begin{figure}
\includegraphics[width=0.96\columnwidth]{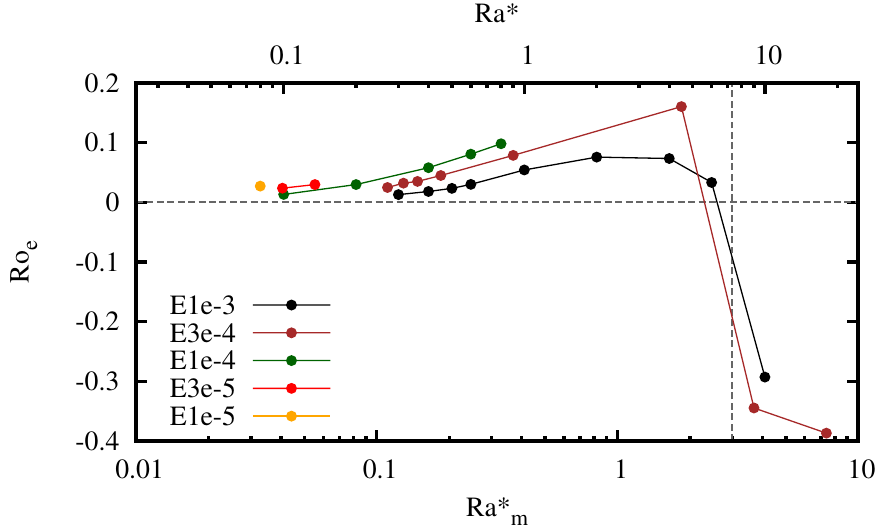}
\centering
\caption{Surface equatorial jet amplitude as a function of $Ra_m^\star$ for
several Ekman numbers.}
\label{rozonq0}
\end{figure}

\revb{We start with discussing reference cases for homogeneous outer boundary heat flux.} \revb{Scanning different Ekman and Rayleigh numbers, we find the two distinct zonal flow regimes identified in previous studies
\citep[e.g.][]{Aurnou2007,Gastine2013}}. \revb{Fig.~\ref{rozonq0} shows how the non-dimensional equatorial zonal flow velocity, $Ro_e = \overline{u}_\phi(r_o, \pi/2)/D \Omega$ changes with Ekman and Rayleigh number. }
 When Coriolis
forces dominate \revb{over buoyancy} (i.e.~$Ra_m^\star < 3 $) convection tends to assume a
$z$-independent geostrophic structure in the form of \revb{tilted} convective columns
aligned with the rotation axis. \revb{As discussed above, the consistent tilt leads to Reynolds stress that drives the typical prograde equatorial jet.} The equatorial jet amplitude increases \revb{consistently with
\rev{$Ra_m^\star$} until the jet changes direction at $Ra_m^\star < 3 $ where the angular momentum is homogenised by turbulent convection. This is reflected by the negative and large value of $Ro_e$ when $Ra_m^\star > 3.0$. 
 Numerically limitations prevent us from reaching the inertia-dominated regime for smaller Ekman numbers.}

\revb{\citet{Aurnou2007,Gastine2013b} found the transition at a slightly smaller $Ra_m^\star \approx 1$. We attribute the discrepancy to different boundary conditions and heating modes used in these papers. However, the reversal of zonal flows still occurs at $Ra_m^\star \approx O(1)$ indicating the universality of this process. }

%Note, whereas the transitional value of $Ra_m^\star$ was found to be $%\sim 1$ for a series of 
%models %with fixed entropy/temperature boundary conditions 
%driven from the inner boundary \citep{Aurnou2007,Gastine2013b}, 
%we find for the here applied fixed flux boundary conditions and internal heating setup a \revb{slightly higher} $Ra_m^\star \approx 3.0$. However, the reversal of zonal flows still occurs at $Ra_m^\star \approx O(1)$ 
%indicating the universality of this process.

%When $Ra_m^\star > 3.0$, buoyancy becomes a first-order
%contribution to the force balance and the flow looses its coherent columnar
%structure. The total angular momentum is then \revb{homogenised} by the 3D turbulent
%convective motions \citep{Gilman79, Aurnou2007}. 
%Because of the larger
%Rayleigh numbers, this buoyancy-dominated regime is
%numerically very demanding and we could only afford to reach it for the two
%largest Ekman numbers explored here. \revb{If} the total angular momentum is homogeneously
%distributed in the fluid shell, the zonal flow is retrograde
%(prograde) at the outer (inner) boundary due to the larger (smaller) distance
%to the rotation axis (fig.~\ref{rozonq0}). 

\begin{figure}
\includegraphics[width=.96\columnwidth]{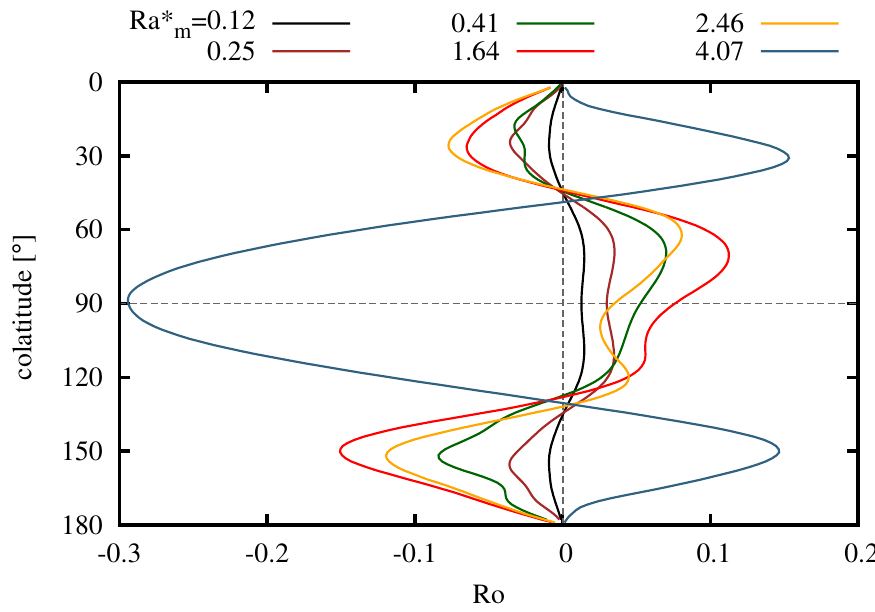}
\caption{Surface zonal flow as function of colatitude for several Rayleigh
numbers at an Ekman number of $E=10^{-3}$.}
\label{zfg0E1e3}
\end{figure}

Fig.~\ref{zfg0E1e3} shows the time averaged surface zonal flow profiles for
increasing Rayleigh numbers at a fixed Ekman number of $E=10^{-3}$. 
\revb{At larger Rayleigh numbers, but before} the transition around \revb{$Ra_m^\star\approx 3.0$}, the zonal flow profiles show significant 
asymmetry with respect to the equator. This can be linked to
the emergence of a \revb{global equatorial temperature asymmetry} which \revb{drives an equatorial antisymmetric thermal wind structure} (eq.~\ref{eqtw}). \revb{This scenario was first reported by
\citet{Landeau2011} for simulations using rigid flow boundary conditions in order to model dynamos of terrestrial planets rather than 
the stress free conditions employed here. 
The symmetry breaking is promoted by volumetric heating and fixed flux thermal outer boundary conditions \citep{Cao2014} but seems independent of the mechanical boundary conditions.} 
\revb{\citet{Cao2014} report that the asymmetry is further promoted by the  $Y_{20}$-shaped anomaly of the outer boundary heat flux when more heat is allowed to escape from the equatorial region, i.e. $q^\star < 0$. }

%In relation to the present study, \citet{Cao2014} reported \revb{that an (equatorial symmetric) $Y_{20}$-shaped heat flux boundary anomaly
%has a direct influence on the EAA-mode.} \revb{The study, using rigid walls rather than free-slip, reports} either a suppression or an %enhancement of equatorial antisymmetric flows depending
%on the sign of the $Y_{20}$ heat flux anomaly.
%When allowing more heat to escape through the equatorial region, the equatorial
%symmetry is more easily broken and a persistent temperature dichotomy
%evolves where either the northern or the southern hemisphere is hotter than
%its counterpart. Strong ageostrophic thermal winds are once more the consequence.

\section{Inhomogeneous outer boundary heat flux}
\label{sec:resultsinhom}

\subsection{Thermal Winds and Zonal flows}
\label{sec:TW_ZF}
\begin{figure}[ht]
\centering
\includegraphics[width=0.96\columnwidth]{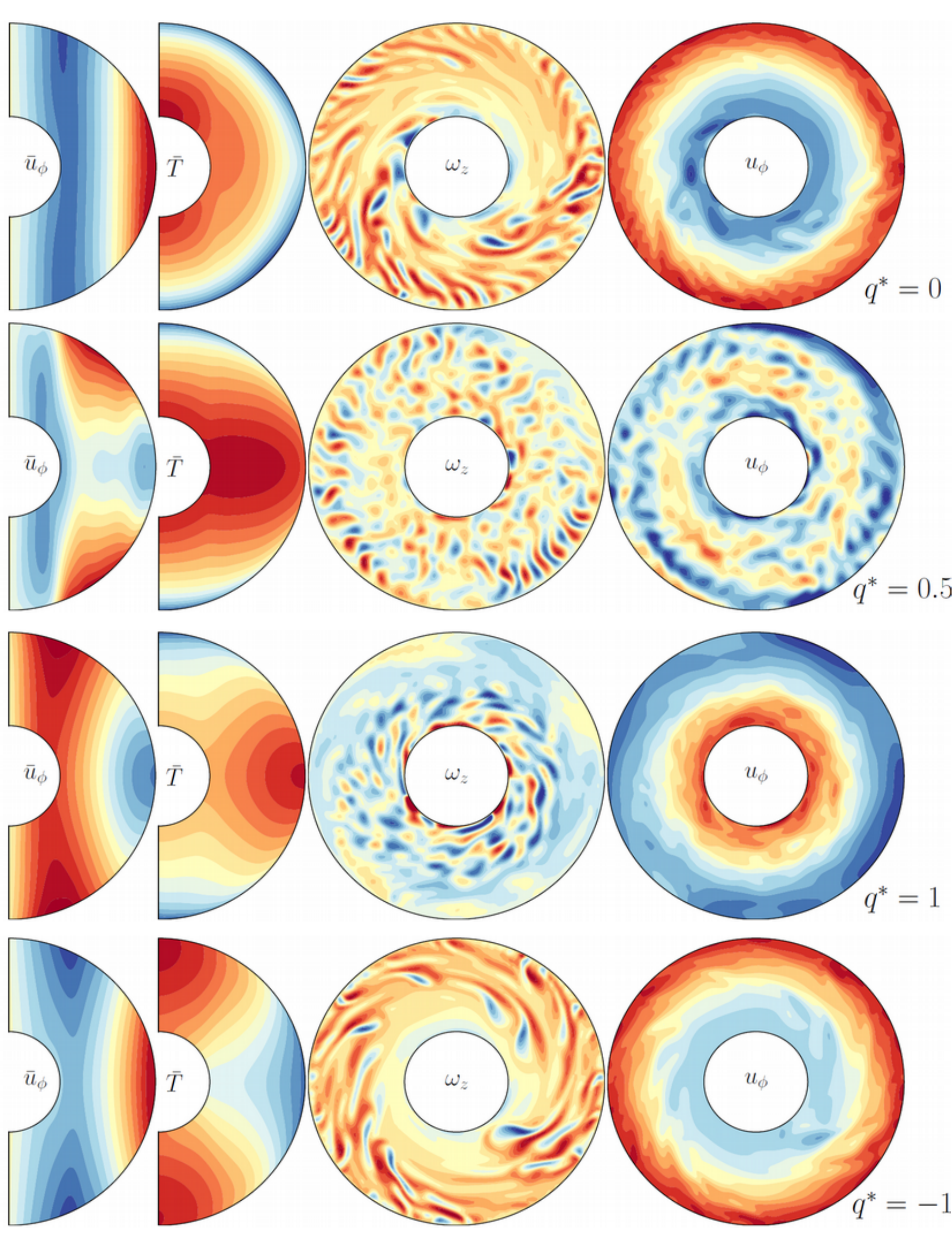}
\caption{The two first columns show time and azimuthally averaged zonal flow
and temperature. The two last columns show snapshots of equatorial cuts of
$z$-vorticity and azimuthal flow.}
\label{flowvarq}
\end{figure}

\begin{figure*}
\centering
\includegraphics[width=0.75\columnwidth]{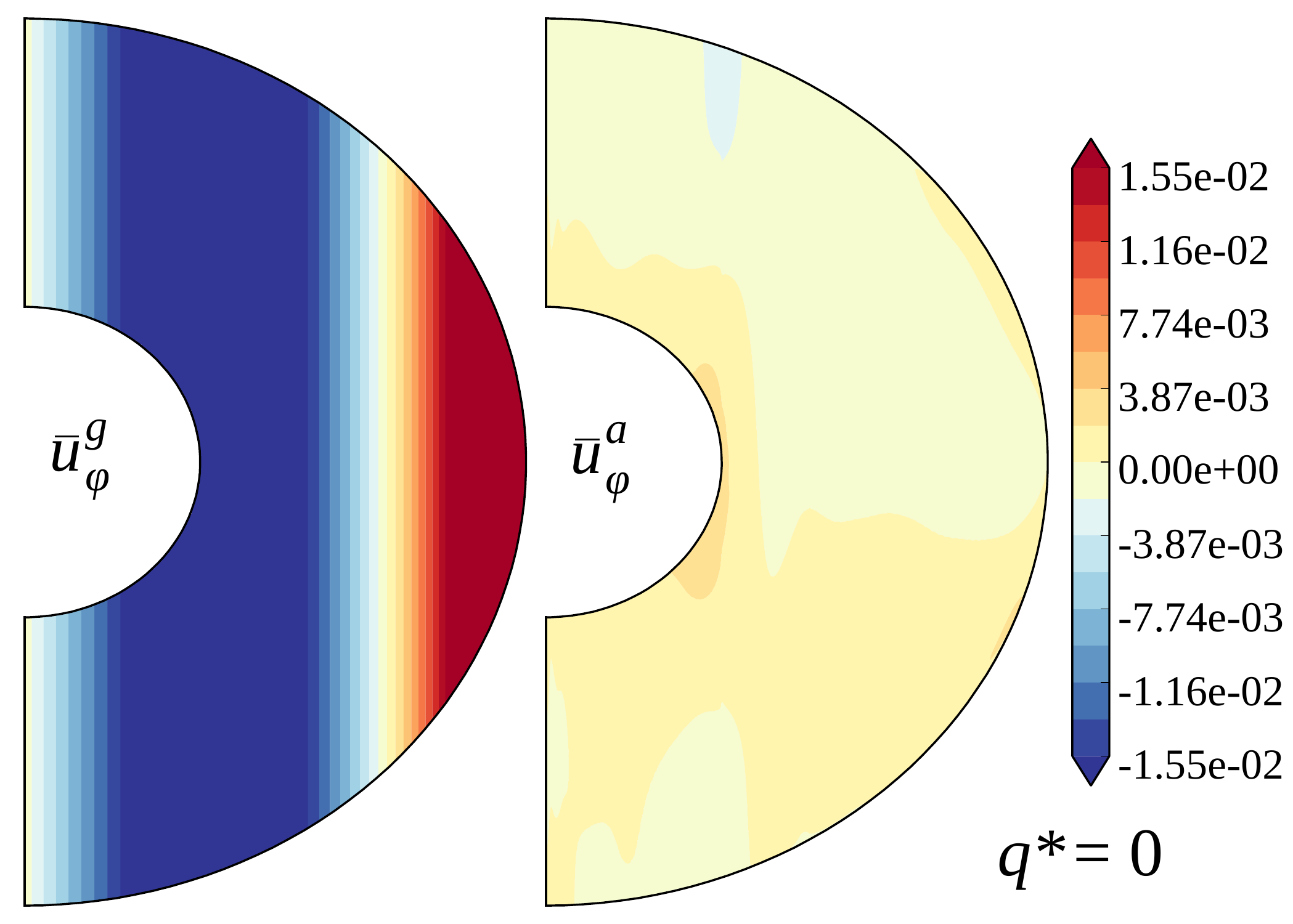}
\includegraphics[width=0.75\columnwidth]{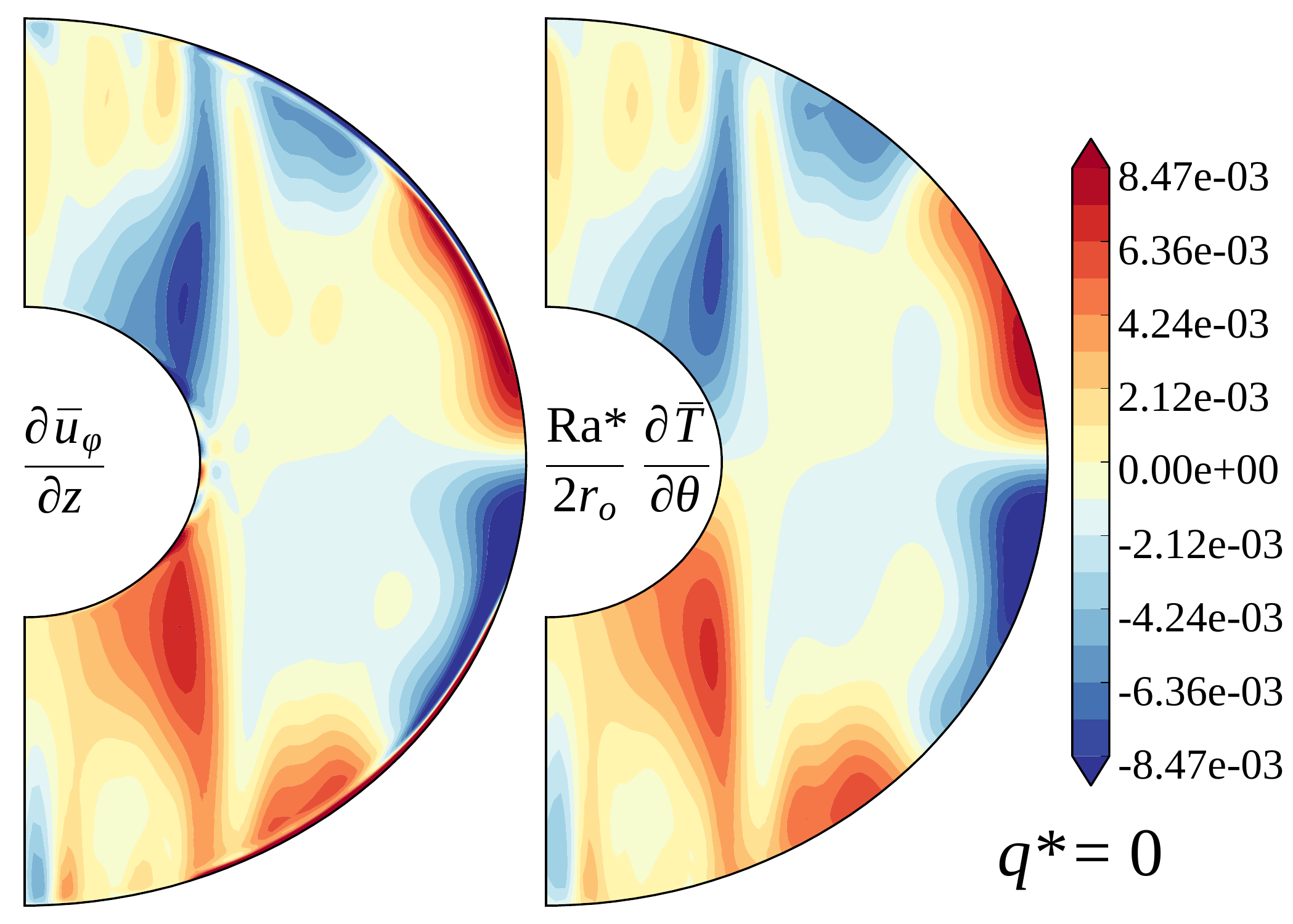}
\includegraphics[width=0.75\columnwidth]{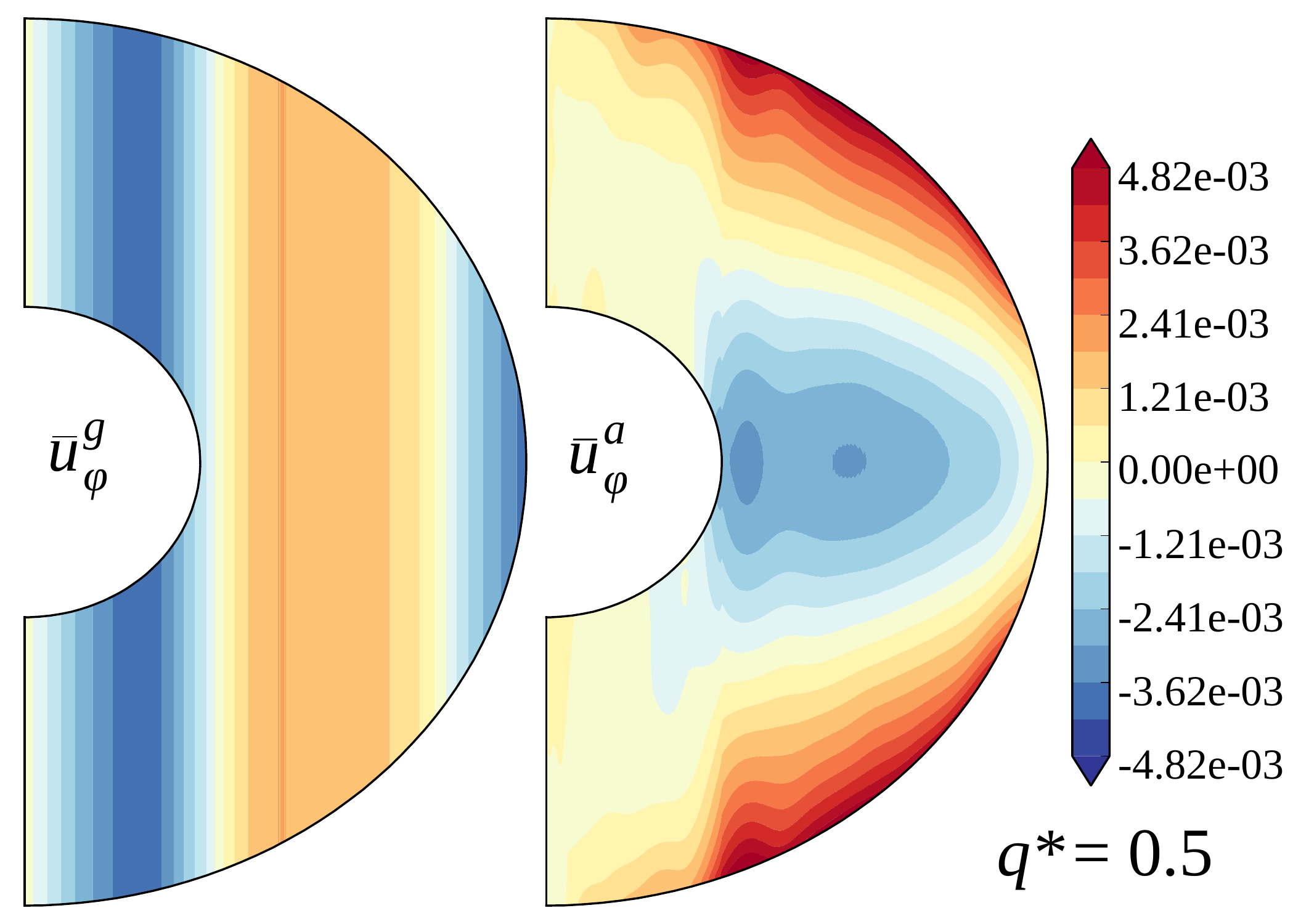}
\includegraphics[width=0.75\columnwidth]{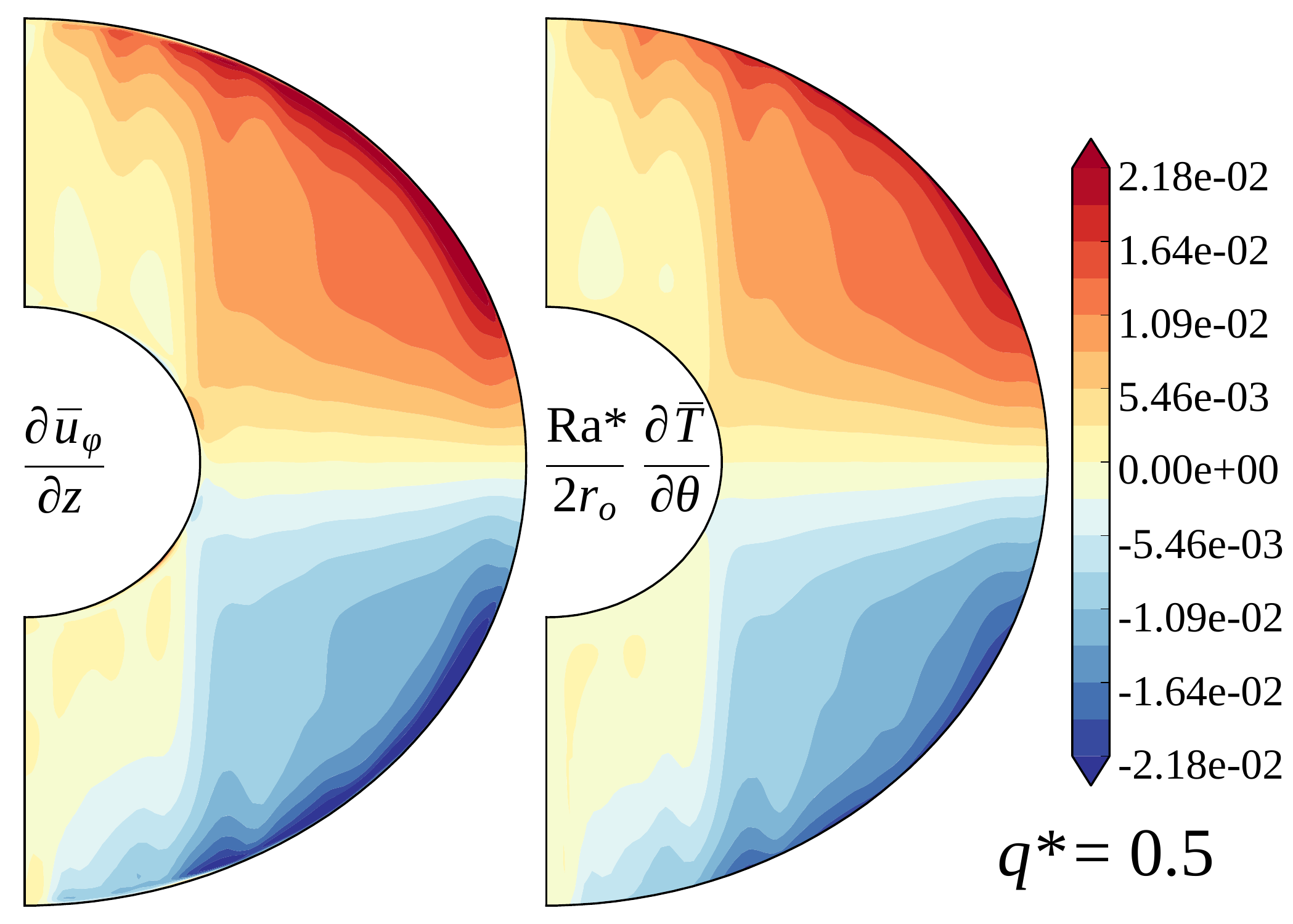}
\includegraphics[width=0.75\columnwidth]{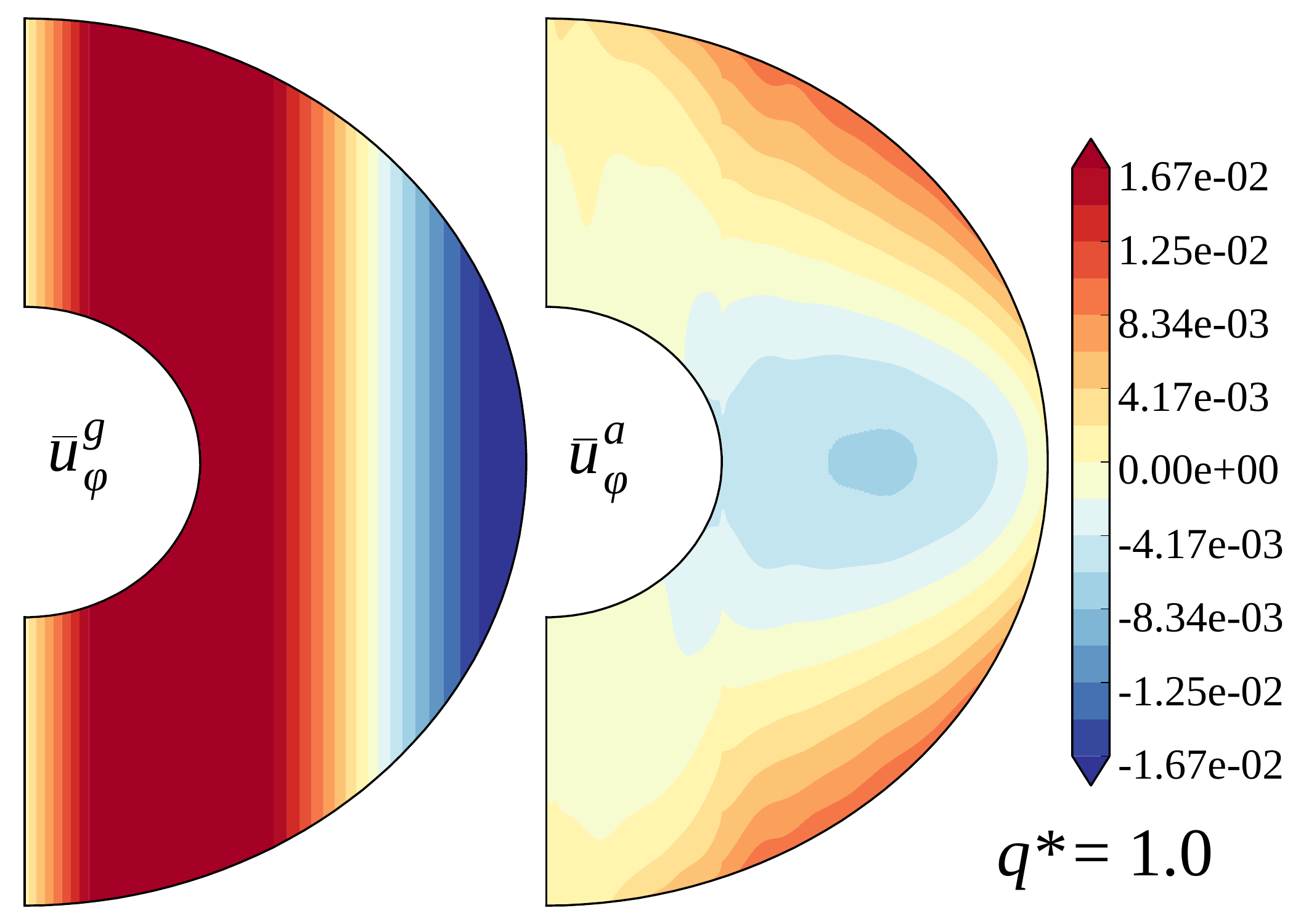}
\includegraphics[width=0.75\columnwidth]{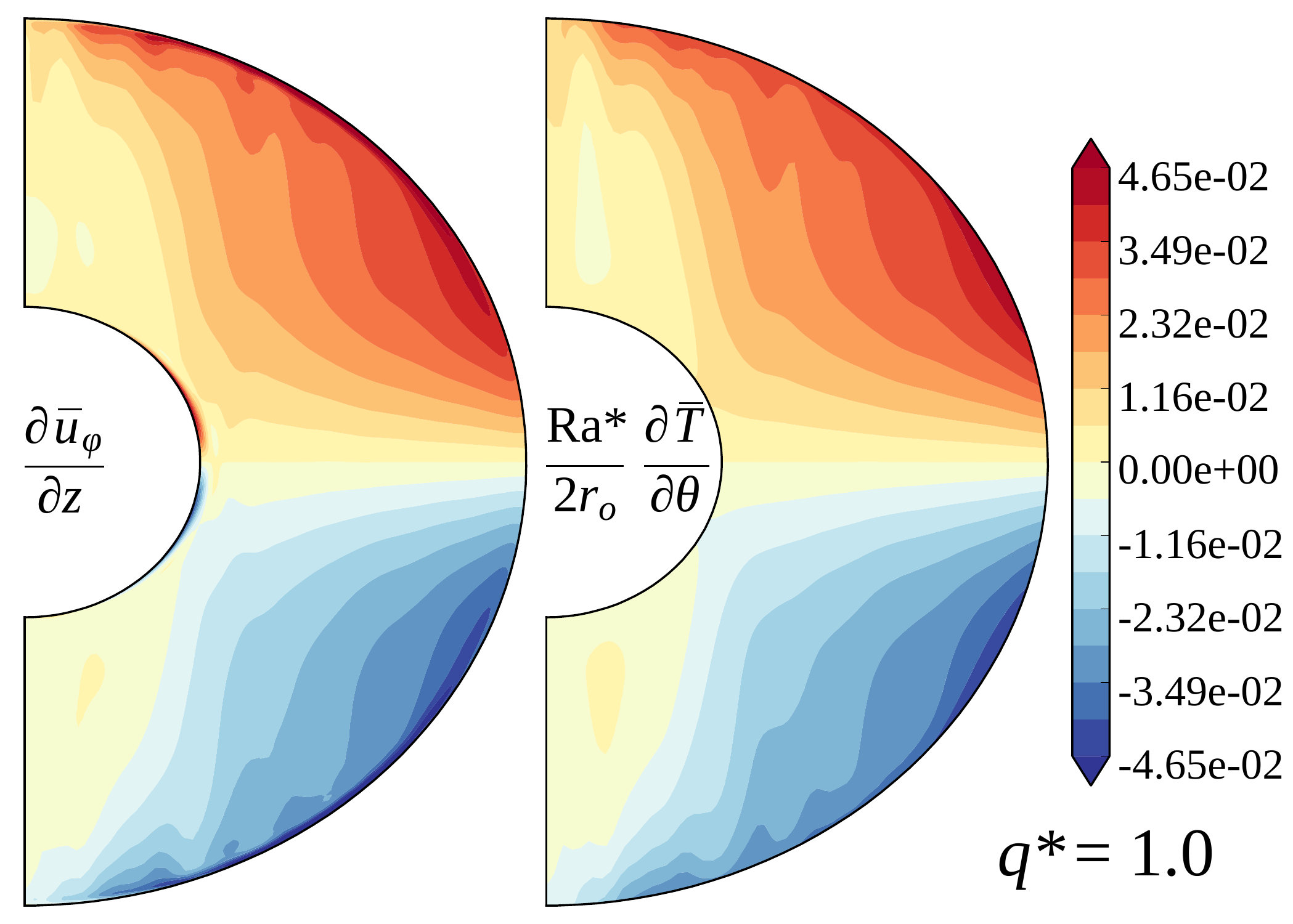}
\includegraphics[width=0.75\columnwidth]{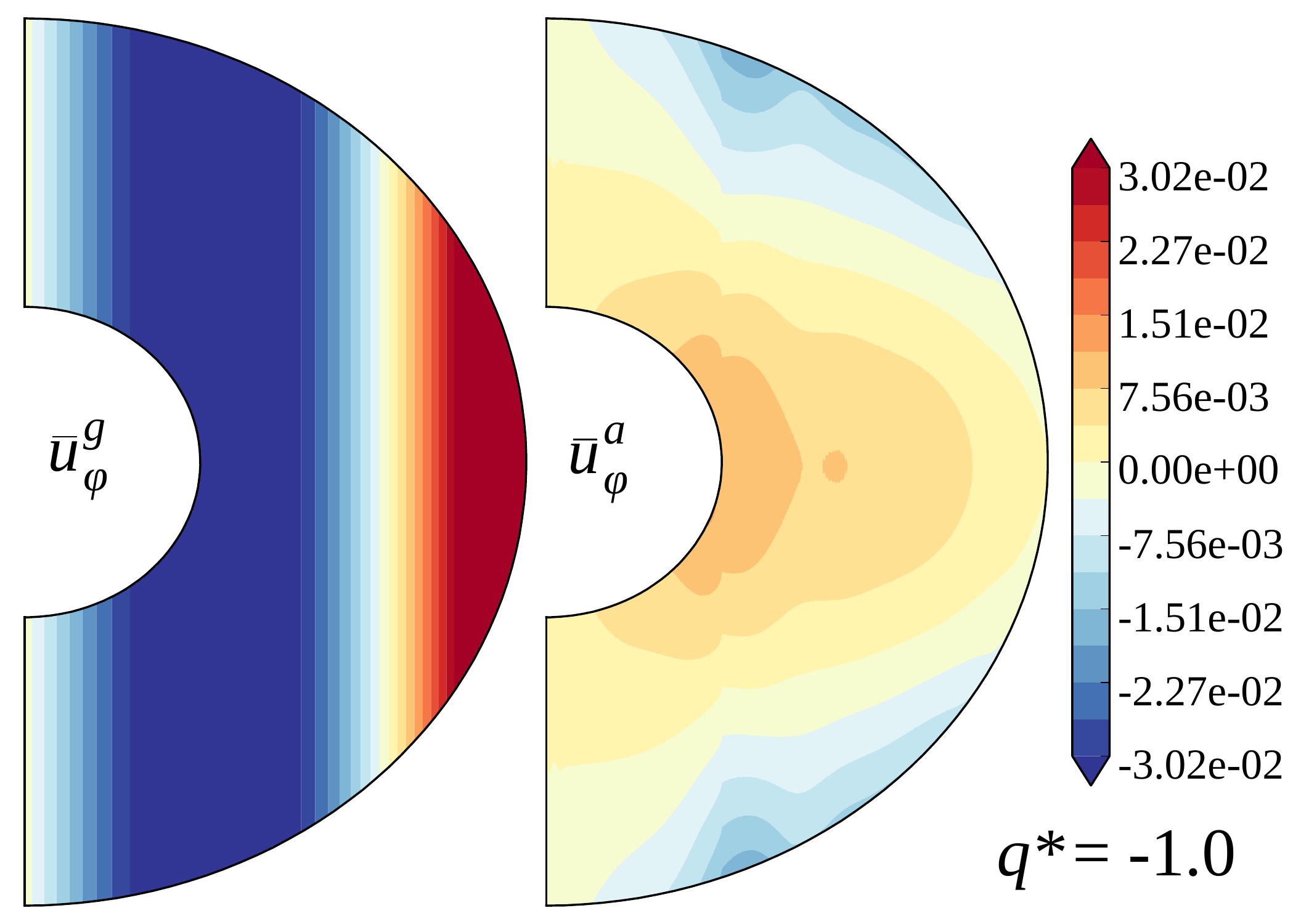}
\includegraphics[width=0.75\columnwidth]{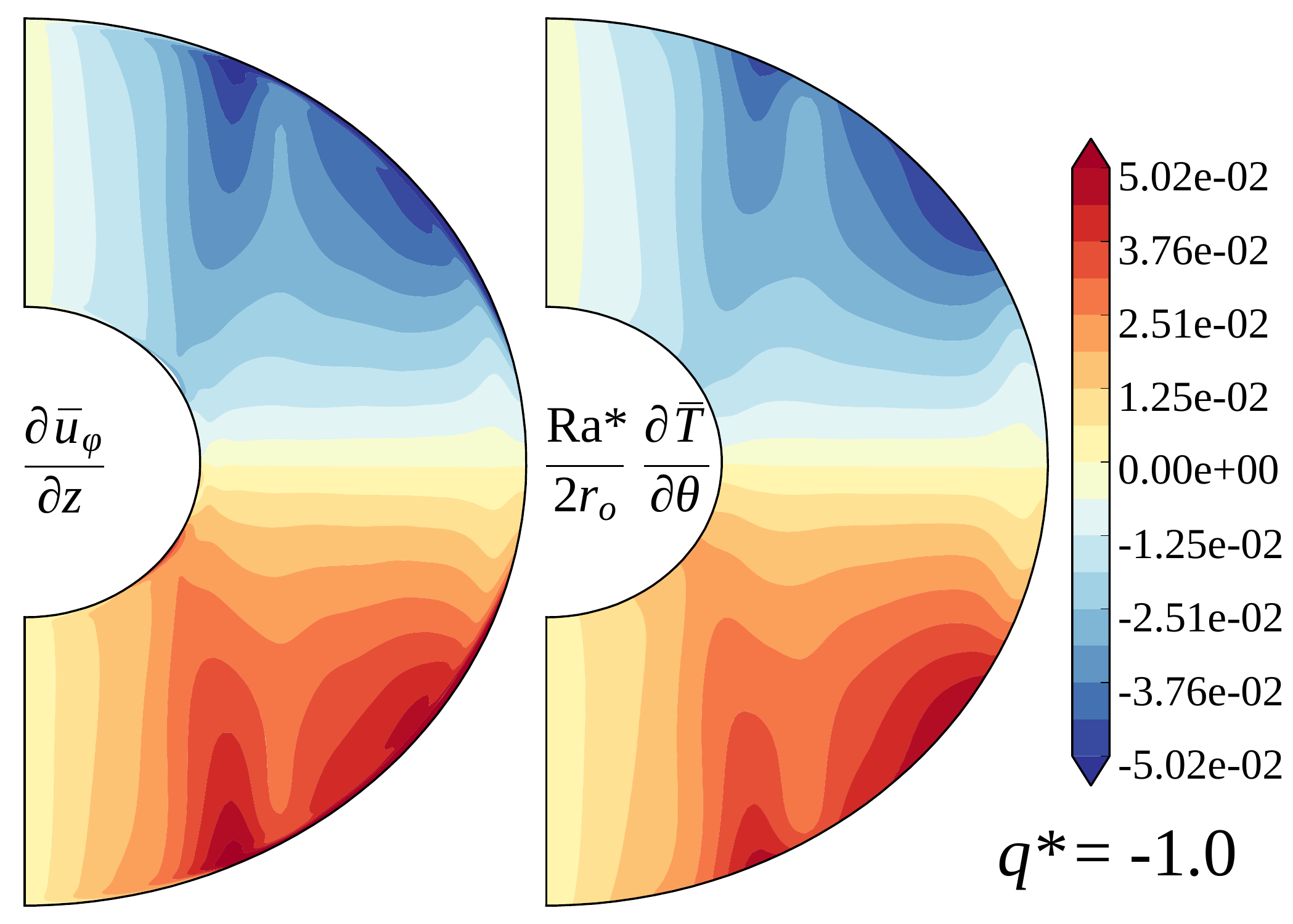}
\caption{First and second columns show the geostrophic and ageostrophic part
of the time-averaged zonal flow. The third and fourth column show the
thermal wind balance. For some cases the contour
levels in the plots of the geostrophic zonal flows are truncated in order to highlight the ageostrophic
contribution since both share the same contour levels.}
\label{thewivarq}
\end{figure*}

To investigate the \revb{interplay} between geostrophic and ageostrophic zonal
flows, we \revb{choose} a reference case \revb{with homogeneous outer boundary heat flux and a moderate Rayleigh number} where the zonal flow structure is
strongly \revb{geostrophic and thermal winds play a minor role.} 
\revb{W}e first consider a numerical model with $E=10^{-4}$ and
$Ra^\star=0.2$ ($Ra^\star_m=0.082$). \revb{For this case} this model yields a prograde
equatorial zonal flow maintained by Reynolds stresses. 
Fig.~\ref{flowvarq} compares time and azimuthal average zonal flow and
temperature for heat flux perturbation amplitudes $q^\star=0.5,1,-1$
with the reference case $q^\star=0$.
The equatorial slices of $z$-vorticity and azimuthal flow illustrate
the \revb{instantaneous} structure.
Fig.~\ref{thewivarq} shows the geostrophic and ageostrophic contributions
of the time averaged zonal flow. The last two
columns of this figure demonstrate that the ageostrophic part is
indeed thermal wind related by illustrating the
good agreement between left- and right-hand-side of the
thermal wind balance eq.~(\ref{eqtw}).

In the $q^\star=0$ reference case (top row of figs.~\ref{flowvarq}, \ref{thewivarq})
the time-averaged zonal flow shows a strongly geostrophic structure with a
prograde (retrograde) zonal flow at large (smaller) distances from the rotation axis \revb{$s$}.
The mostly radial isotherms (same figure, second column) confirm that latitudinal
temperature gradients are too small to drive significant ageostrophic thermal winds.
The snapshot of the equatorial $z$-vorticity shows mainly positive
values and illustrates the consistent prograde tilt of the
geostrophic columns.

The second rows of figs.~\ref{flowvarq}, \ref{thewivarq} show the
solution for a moderate anomaly  $q^\star=0.5$.
At this value the heat flux at the equator (the pole) is half (1.5 times)
the mean flux. Consequently the poles are cooled more efficiently by
convection and a positive (negative) latitudinal temperature gradient
is established in the northern (southern) hemisphere.
The resulting thermal winds are retrograde in the equatorial
region and prograde towards both poles.
The geostropic zonal flow has also changed fundamentally and is now retrograde
at large $s$, prograde at intermediate $s$ and again retrograde at small
$s$. This seems roughly consistent with a change in the tilt of convective
features illustrated in column three and four of fig.~\ref{flowvarq}.
Some retrograde tilted features can now be identified at large $s$ where
the new retrograde zonal wind appeared.

\revb{If the perturbation amplitude is further increased} to $q^\star=1$ thermal
winds gain in amplitude while preserving a very similar structure.
The geostrophic \revb{flows}, however, have now reversed completely with a
strong retrograde jet \revb{at the equator} and wide prograde jets at
smaller \revb{$s$}.
Convective \revb{flows} (fig.~\ref{flowvarq}, third row) are now rather weak \revb{closer to the outer boundary} where the
suppressed heat flux leaves a \revb{neutrally stratified} hot equatorial region. \revb{Thermal winds are significantly stronger than for $\qs=0.5$}. \revb{The columns} are now tilted  dominantly in retrograde direction
(see z-vorticity and $u_\phi$ snapshots in column 3 and 4 of figure \ref{flowvarq}). 
\revb{As we will further discuss below, this leads to a reversed direction of Reynolds stresses and consequently to a geostrophic zonal flow with retrograde equatorial and prograde inner jet. Whereas the influence of an increased $\qs$ on the thermal wind is as expected, the influence on the tilt and thus the geostrophic zonal flow direction comes as a surprise. Since the Reynolds stress driven geostrophic zonal flow component is once more clearly dominant, the relative vertical variation is weaker than for the $\qs=0.5$ model.}
%This implies that the hot equator is aligned with retrograde zonal flows and the prograde jets with lower temperatures in higher latitudes.
%Note, that the relative vertical variation is weaker than for the $\qs=0.5$-case, but still entirely determined by the latitudinal temperature gradient (fig.~\ref{thewivarq}, third column). However, the geostrophic part of the zonal flow is re-amplified by the inverse Reynolds-stresses. This is indeed unexpected, as an increase of $\qs$ should only lead to larger ageostrophic flows. However, what we find is an drastic increase (and reversal) of the geostrophic zonal flow system.}

We also tested the case $q^\star=-1$ where the heat flux is enhanced in the
\rev{equatorial region} but reduced \rev{towards the poles}. The last
columns in figs.~\ref{flowvarq} and \ref{thewivarq} show that the
thermal wind is as strong as in the $q^\star=1$ simulations but, as
expected, simply has reversed its sign. The geostrophic zonal flows
increase in amplitude but retain their structure. Likewise, convective
features have a similar distribution and \revb{the same tilt direction} as in the $q^\star=0$
reference case. \revb{However, the Reynolds stresses are more efficient as demonstrated by the non-axisymmetric flow correlation listed in tab.~\ref{tabflow} and therefore drives stronger geostrophic zonal flows.}

\begin{figure}
\centering
\includegraphics[width=0.96\columnwidth]{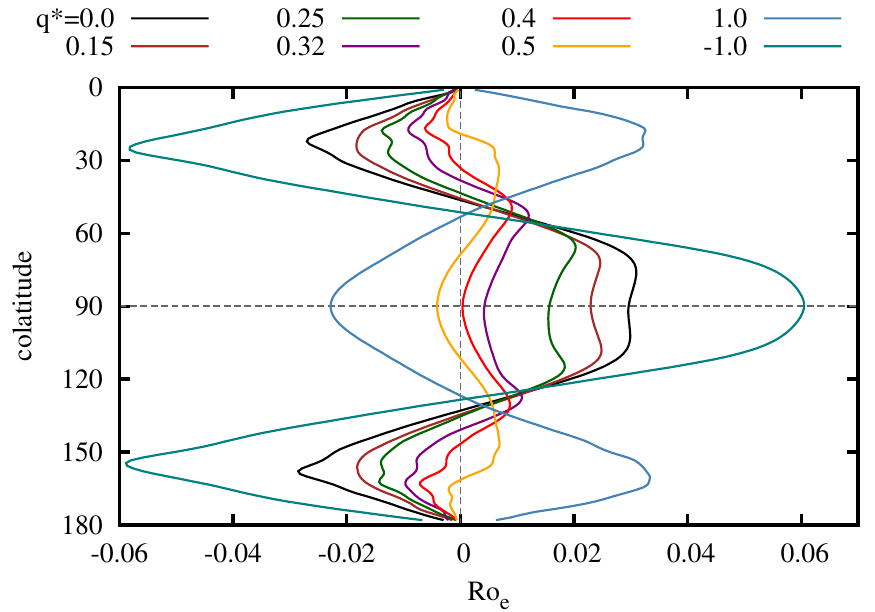}
\caption{Surface zonal flow structure for several perturbations amplitude
$q^\star$. From $q^\star\simeq 0.5$ the thermal forcing reverts the equatorial
jet from prograde to retrograde. Parameters: $E=10^{-4}, \,Ra^\star=0.2,\,
Pr=1$.}
\label{rozonvarq}
\end{figure}

Fig.~\ref{rozonvarq} shows the \revb{changes of the} surface zonal flow patterns \revb{when}
$q^\star$ \revb{is increased}. 
\revb{If the non-homogeneous outer boundary condition would only drive thermal wind but leave the geostrophic zonal flow largely unaffected, the surface flow profile would remain unchanged at the equator and become more prograde at mid to high latitudes. The high latitude retrograde jets would be weakened or reversed, while the low-latitude prograde jet would be intensified except directly at the equator. }

\revb{However, due to the fact that the geostrophic winds also change drastically, the behaviour appears very different. The reduction and ultimately reversal of Reynolds stress driven geostrophic zonal flows starts at the equator as $\qs$ increases and propagates inwards. The equatorial jet therefore first develops a dimple or minimum around the equator before it turns retrograde at even higher $\qs$. The dimple deepens with increasing $\qs$ and is reminiscent of a similar feature in Jupiter's main prograde jet.}

%The strength of the dimple relative to the flanking jets appears to be function of $\qs$ (fig.~\ref{rozonvarq}) and is reminiscent to the large dimple observed in Jupiter's main equatorial band.}

At $q^\star=0.4$ the equatorial zonal flow
practically vanishes \revb{at the outer boundary} while stronger perturbations promote increasingly retrograde
equatorial jets. The respective flow profiles are then similar\revb{ly shaped to} the ones observed
for Neptune and Uranus \citep[e.g.]{Aurnou2007}.
\revb{For an} anomaly with negative sign ($q^\star=-1$) the structure is similar to 
the homogeneous ($q^\star=0$)-model but with twice the amplitude.

%\revb{When increasing $\qs$ the overall zonal flow amplitude decreases}
%Applying the heat flux
%anomaly with weak amplitude gradually decreases the amplitude of the surface zonal flow.
%The moderately forced cases ($0.15<q^\star<0.4$) feature an increasing dimple of
%the equatorial jet peak, which is reminiscent of the large dimple observed
%in Jupiter's main equatorial band. 

\subsection{Force Balance}
\label{ssforcebal}
\begin{table}[t]
\centering
\renewcommand{\arraystretch}{1.4} % Default value: 1
\begin{tabular}{c|ccc|c}
q & 0 & 0.5 & 1 & -1 \\
\hline
$Re_{\overline{\phi}}$& 364.43& 58.84  & 335.22 & 619.61 \\
$Re_{\overline{s}}$   & 0.878  & 0.857  & 1.751 & 0.885 \\
%\hline
$Re_{\phi^\prime}$  & 78.50 & 67.81 & 81.72 & 96.11 \\
$Re_{s^\prime}$     & 56.12 & 63.08 & 56.73 & 55.99 \\
%\hline
$\cal{RS}$ & 1054.1 & 65.59 & 731.7 & 1569.6 \\
$\cal{AD}$ & 10.1 & 6.74 & 113.1 & 8.13 \\
\end{tabular}
\caption{Several flow properties for the four study cases with $q=0$, 0.5, 1, and -1. \rev{The individual Reynolds numbers are calculated for the axisymmetric (overbar) and non-axisymmetric (primed) flows along azimuth ($\phi$) and cylindrical radial (s).}}
\label{tabflow}
\end{table}

\begin{figure*}
\centering
\includegraphics[draft=false,width=0.8\textwidth]{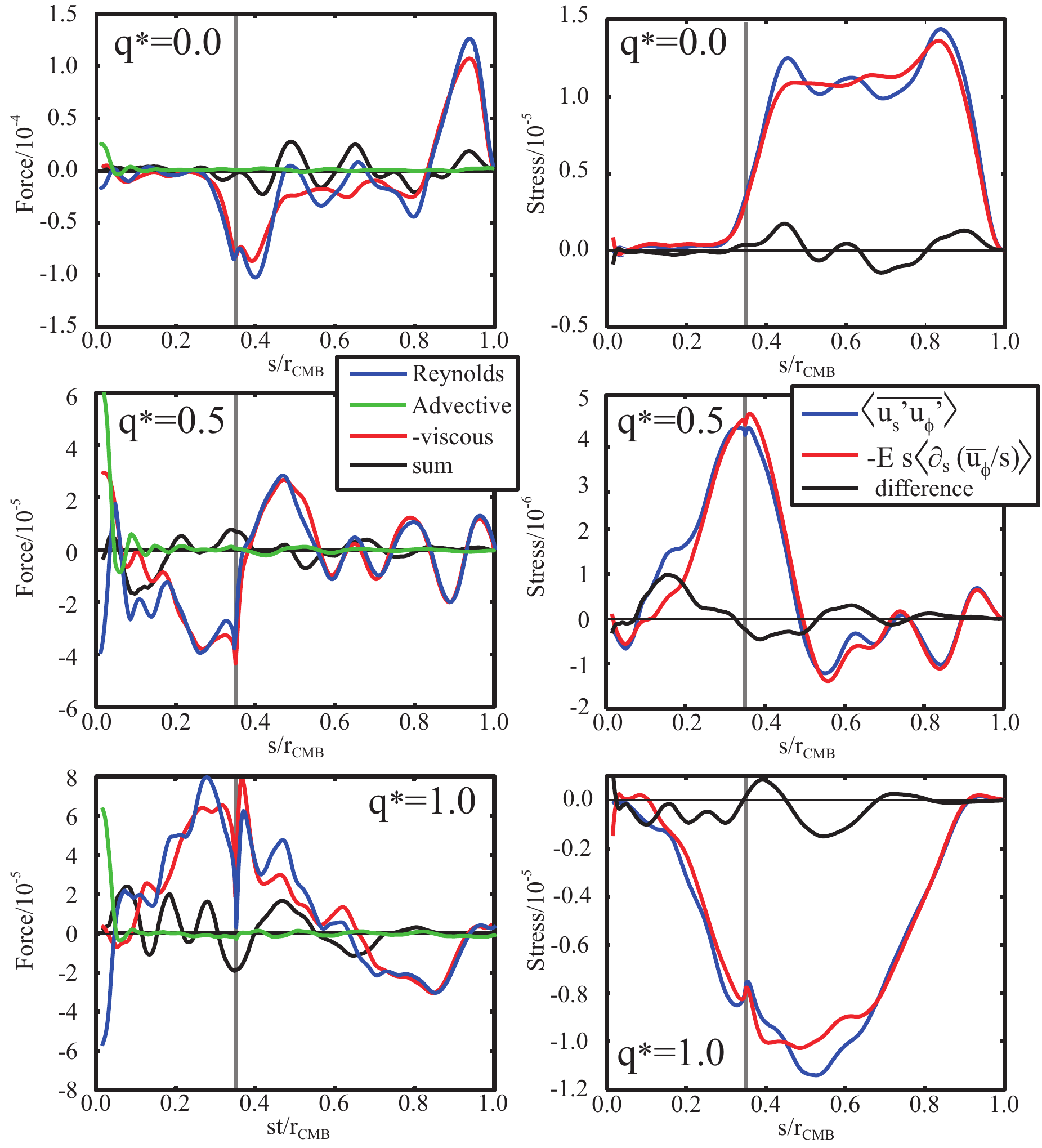}
\caption{The left column shows the z-averaged \rev{force due to} Reynolds stress \rev{(blue)},
the advective force \rev{(green}), minus the viscous force \rev{(red)}, and the sum of the two forces \rev{(black)} for three
of the four cases depicted in \figref{flowvarq}.
The right column shows respective z-averaged contributions to the
balance between the Reynolds stress \rev{(blue)} and the viscous \rev{stress (red)} according to \eqnref{UT}. The vertical black lines denote the tangent cylinder.}
\label{Int}
\end{figure*}

\begin{figure}
\centering
\includegraphics[draft=false,width=0.8\columnwidth]{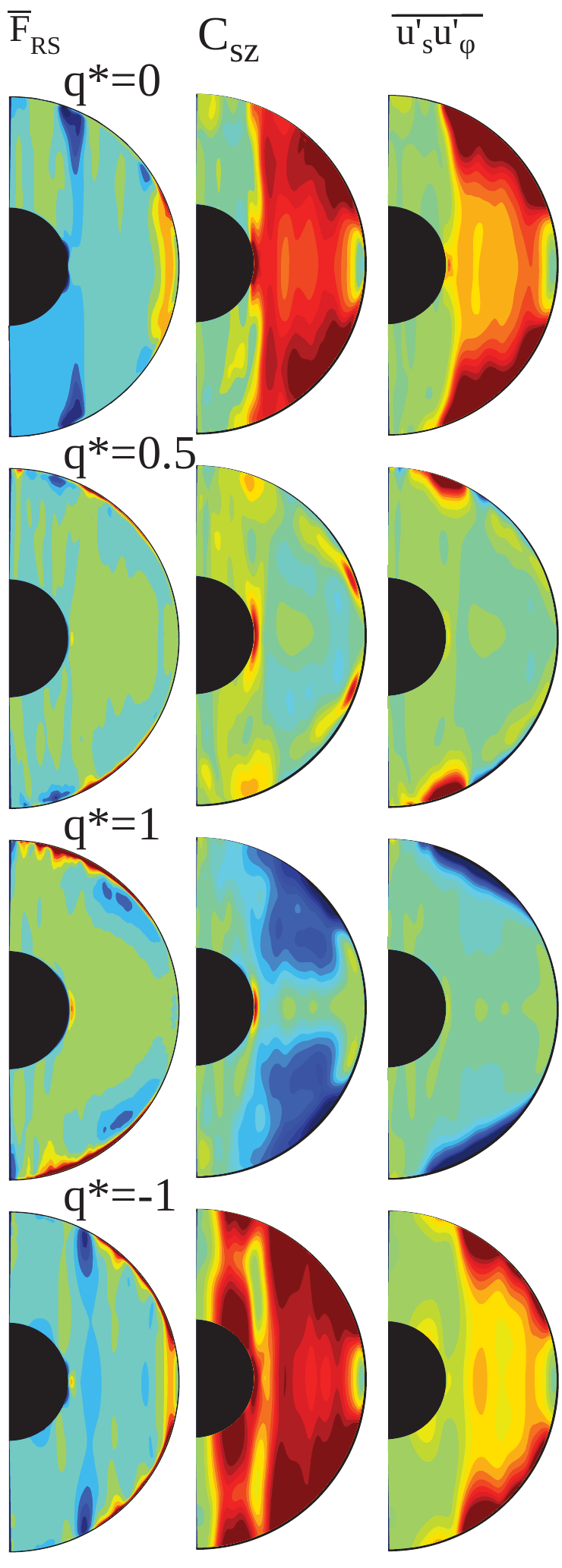}
\caption{Time averaged axisymmetric azimuthal maps of Reynolds stress \rev{convergence},
correlation $\corr$ and \rev{Reynolds stress} ($\overline{u_\phi^\prime u_s^\prime}$) for the four
selected cases at different $\qs$ values already depicted in \figref{flowvarq}.
Contours for the first and second columns have been chosen to
individually highlight the structure. The correlation shown
in column two varies between $-0.5$ and $0.5$.}
\label{MerForceBallance}
\end{figure}

\Figref{Int} shows the $z$ and time averaged \rev{zonal force balance due to Reynolds stress
convergence\revb{, advective force}
and viscous force} (left column) for the cases at $Ra^\star=0.2$ and $E=10^{-4}$ already discussed in
\rev{\secref{sec:TW_ZF}}. \revb{As suggested in tab.~\ref{tabflow} the advective force
due to meridional circulation is negligible for all cases.}
The two dominant terms in the \rev{stress} balance
(\ref{UT}) \revb{are displayed in the right column}. Results for $\qs=-1$ are not shown since they are
very similar to that for $\qs=0$.
Note that the Reynolds stress and viscous contributions do not balance
perfectly mostly because we have not averaged long enough in time.
\Figref{MerForceBallance} compares the time averaged zonal Reynolds stress \rev{convergence},
correlation $\corr$, and the azimuthally averaged nonlinear product
$\overline{u_s^\prime u_\phi^\prime}$\rev{, e.g. the Reynolds stress.} The correlation coefficient $\corr$ is defined as
\begin{equation}
 \corr (s,z) = \frac{\overline {u_s^\prime u_\phi^\prime}}{\left(\overline{{u_s^\prime}^2} \overline{{u_\phi^\prime}^2}\right)^{1/2}} \ .
\end{equation}

For a homogeneous \rev{outer boundary} heat flux, positive \revb{(negative)} Reynolds stress \rev{convergence} close to the outer
boundary \revb{(tangent cylinder)}  is
responsible for driving the prograde equatorial and retrograde inner
zonal jets (first columns in \figref{Int} and \figref{MerForceBallance}).
The correlation $\corr$ is strongly positive outside the tangent cylinder
due to the generally prograde tilt of the convective features here
(second column in \figref{MerForceBallance}).
\revb{T}he region close to the outer boundary contributes more
to $\overline{u_s^\prime u_\phi^\prime}$ and thus to the Reynolds stress
because of higher $u_\phi^\prime$ and $u_s^\prime$ amplitudes
(column three in \figref{MerForceBallance}).
The $z$-average $\geos{\overline{u_s^\prime u_\phi^\prime}}$ (second column in fig.~\ref{Int})
shows a pronounced positive hump between the tangent cylinder and \revb{the outer boundary equator}.
Steep gradients close to these two boundaries translate into the two dominating
Reynolds stress \rev{convergence} features.

At $q^\star=0.5$ the correlation and thus the Reynolds stress is very low.
The positive hump in $\geos{\overline{u_s^\prime u_\phi^\prime}}$ is now located at
the tangent cylinder, resulting in prograde \revb{(retrograde}) Reynolds stress \rev{convergence} attached to the
inside \revb{(outside)} of the TC. \revb{\Figref{MerForceBallance} illustrates that the reason is the strong concentration of non-axisymmetric flow components close to where the TC touches the outer boundary.}

At $\qs=1$ the correlation $\corr(s,z)$ is once again generally strong outside
the tangent cylinder except for a region around the equatorial plane. The sign is now
negative, however, due to the retrograde tilt of non-axisymmetric
convective features. 

Once more, $\overline{u_s^\prime u_\phi^\prime}$ and thus the Reynolds stress \rev{convergence} 
are strongest closer to the outer boundary and at mid to higher latitudes.
The negative hump in $\geos{\overline{u_s^\prime u_\phi^\prime}}$ reaches inside
the tangent cylinder, leading to a strong positive \revb{force due to} Reynolds stress there.

As already discussed above, the test case for $\qs=-1$ is generally very similar
to the reference case $\qs=0$. However, the positive correlation now also stretches inside
the tangent cylinder while the non-axisymmetric flow is somewhat stronger promoted
at lower latitudes. The combination of both results in a Reynolds stress
pattern similar to that at $\qs=0$.

The primary flow component driven by the boundary inhomogeneity
is the thermal wind. However, the advective force via which the thermal
wind could directly contribute to \revb{drive} geostrophic zonal flows
remains very small for all the cases explored here \tabrefp{tabflow}.
The Reynolds stress and correlation analysis confirm what we
had already anticipated in \secref{flowvarq}: the zonal flow inversion is related to the global change
in the sign of the correlation $\corr$ which in turn is caused by the inversion of the tilt. In the next section we discuss which mechanisms could be responsible for changing the tilt.
Other changes in the non-axisymmetric convective flow are also pronounced
but mostly concern the distribution of Reynolds stress and not its general direction.
\revb{For example, we find an unexpected strong contribution of equatorial antisymmetric, non-axisymmetric convective flows at $\qs>0.5$}.

\section{Changing the tilt}
\label{sec:tilt}

The tilt of convective features and the related zonal flow generation by
Reynolds stresses has been extensively discussed in the literature
and we refer to \citet{Busse2002} and \citet{Takehiro2008} for overviews.
In fast rotating spherical shells convection sets in as
thermal Rossby waves that have the form of geostrophic
columns aligned with the rotation axis \citep{Busse1970}.
These waves drift in azimuthal direction with a velocity that is determined by
the height change of the container. When the height $h$ decreases with the cylindrical
radius $s$, as is the case outside the tangent cylinder, the waves drift in prograde
direction. A tilted or spiralling form in the $s$\revb{-}direction
is assumed when the phase $\Phi_0$ of the wave
depends on $s$. When $\partial \Phi_0 / \partial s >0$ the columns are tilted in
prograde direction.

\rev{\citet{Busse1982} suggested that} the tilting direction of the convective columns is 
controlled by the curvature of the confining walls.
They study a cylindrical annulus with curved upper
and lower caps, a system that captures the essence of the
thermal Rossby waves while reducing the impact of meridional circulation.
Using asymptotic analysis and laboratory experiments they show that convex
caps similar to those in spherical shells yield drifting thermal Rossby waves that
tilt in prograde direction. The tilt is reversed for concave caps.
Their asymptotic analysis confirms
that \revb{the $s$-gradients of} $\Phi_0$ indeed depends directly on the boundary curvature
in the way indicated by the experiments.
A simple argument discussed by \citep{Busse2002} links the phase to the
potential drift speed of the wave. Since $\partial h / \partial s$
increases with $s$ outside the tangent cylinder the wave should
drift faster at larger than at smaller $s$.
The solution which drifts with only one velocity adapts to this
situation by assuming an $s$-dependent phase $\Phi_0(s)$. \rev{The actual mechanism that leads to the tilt was discussed later by \citet{Takehiro2008}}

\citet{Busse1982} and \citet{Busse2002} discuss various issues that can influence the phase relation $\Phi_0(s)$. One example is the onset of a secondary
instability with the same azimuthal wave number as the primary instability but a different radial
dependence. Differences in the heating mode are another alternative first mentioned
by \citet{Busse1982} and later explored in more detail
by \citet{Takehiro2008}. Using a simplified annulus system, he
showed that the typical prograde tilt results when energy is mostly fed into
the system close to the inner boundary but has to travel outward to larger
radii where most of the energy is dissipated. When the locations of
primary energy input and dissipation are reversed, however, the tilt changes to retrograde.
He concludes that the propagation of the Rossby wave in $s$-direction is the key
physical process that controls the tilt direction. \rev{The applicability of this theory
was extended towards spherical systems by \citet{Takehiro2010}.}

\begin{figure}
\includegraphics[width=0.8\columnwidth]{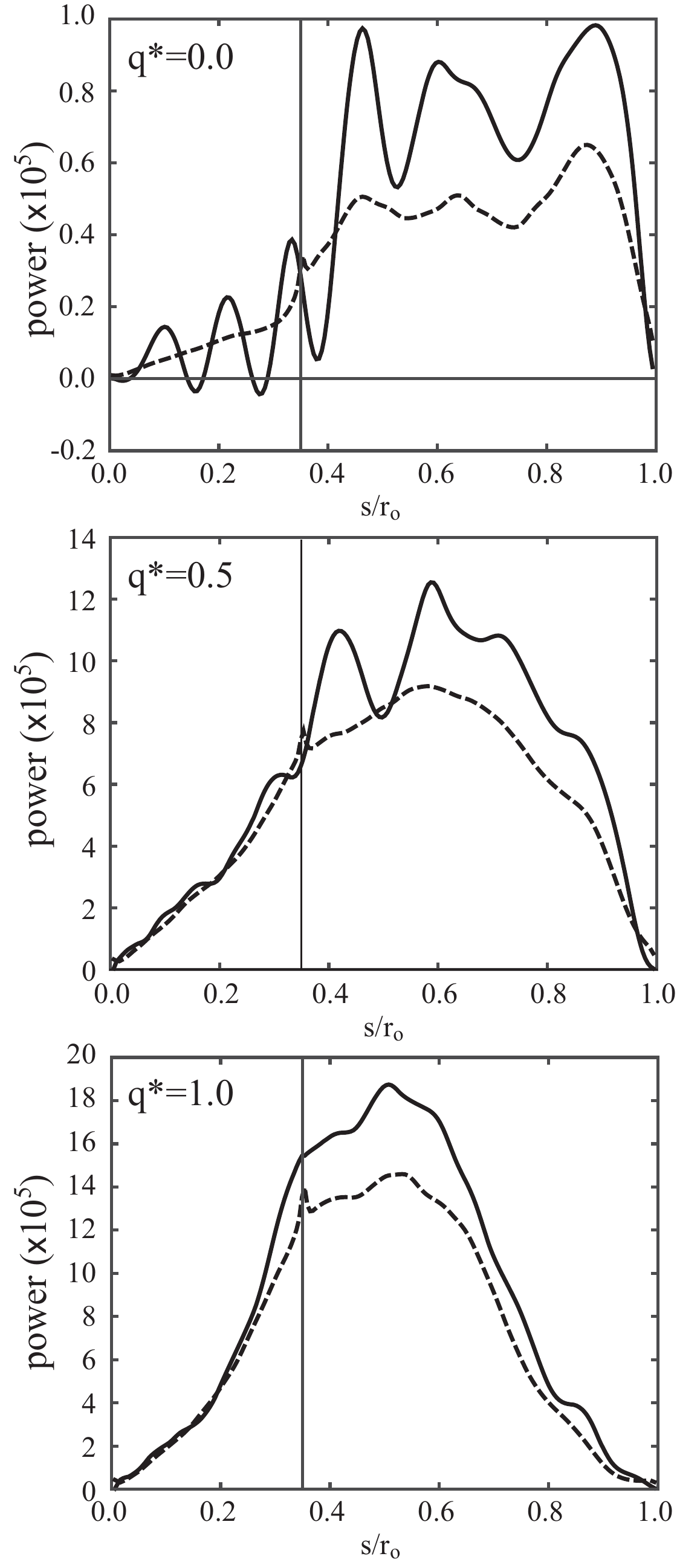}
\centering
\caption{Energy input (solid) and diffusion (dashed) \rev{of non-axisymmetric kinetic energy} integrated over geostrophic
cylinders and averaged over times for the three selected
cases at $Ra^\star=0.2$ and $E=10^{-4}$. The vertical \revb{grey} lines denote the tangent cylinder.}
\label{BuoDiff}
\end{figure}

\Figref{BuoDiff} compares the energy input \rev{$Ra^\star r/r_o u^\prime_r T$}  with
\rev{non-axisymmetric} energy diffusion \rev{$\uvecp\cdot(\nabla^2 \uvec)$} integrated over geostrophic cylinders
and over time for the different cases discussed above.
Both terms result from multiplying the Navier-Stokes equation (\ref{eq:nrj})
with \revb{the non-axisymmetric} velocity $\uvecp$.
When $\qs$ is increased the energy input moves progressively away from the outer boundary
and is clearly concentrated at the tangent cylinder for $q^\star=1.0$.
Energy input and diffusion always have very similar distributions making
it impossible to deduce a potential Rossby wave propagation in $s$\revb{-}direction.
As the zonal flow is reversed consistently with the also reversed Reynolds stress for
$\qs=1.0$, a misalignment of energy input and diffusion should be visible, what is clearly
not the case.  Hence the switch from outward to inward propagation 
on increasing $\qs$ is not explainable with this \revb{rationale}. 

Following the line of arguments by \citep{Busse1982} and \citep{Busse2002} the
tilt change could also be explained when the region inside the tangent
cylinder becomes more important for determining the Rossby wave properties.
 The \revb{inverted} height gradient $\partial h / \partial s$ in this region
 could explain the change in Rossby wave drift and tilt.
However, while \figref{BuoDiff} certainly suggest that the region inside the
tangent cylinder becomes more important on increasing $\qs$, the total energy input is still dominated
by the region outside the tangent cylinder even at $\qs=1$.

Obviously, neither the curvature argument invoked by \citet{Busse1982} nor the Rossby
wave \revb{reasoning} proposed by \citet{Takehiro2008, Takehiro2010}
explain the change in tilt in an obvious way.
This may \revb{not be} surprising since both assume geostrophic
fundamental solutions. While the $\qs=0$ cases may still be more or less compliant
with \rev{the} assumption \rev{of dominant geostrophy}, this changes once the non-homogeneous
boundary condition becomes more influential. The decisively non-geostrophic solutions found at $\qs\approx0.5$ may
require a completely different approach.

It remains to be said that the tilt is generally compatible with
the geostrophic flow directions in the sense that an, for whatever
reason, inverted geostrophic flow pattern would also invert the tilt.

\revb{The simple reason is the twisting that any zonal shear exerts on the
columns}. \citet{Busse2002} \revb{suggest that the tilt direction could be determined by a run-away effect he called the mean flow instability.}
 Suppose, that a system with an undecided columnar tilt \revb{is subjected to a small initial zonal wind shear which could be the result of an instability. The shear most likely tilts the columns, spawns Reynolds stresses and thereby amplifies itself.}
This run-away effect would ultimately stop when viscous and Reynolds stresses
balance.
The effect of the initial zonal wind needs to be strong enough to overcome
any other potentially tilting mechanism discussed above. A small
noise fluctuation would likely not suffice.

\section{Parameter Dependence}
\label{sec:params}
%\revb{INTRODUCE R earlier and include in the table1, and overbar{R}, remove dphi reduce integral to only over s}

\subsection{Anomaly amplitude}
To complete and \revb{generalise} the picture \revb{outlined} above, we investigate several
diagnostic quantities as a function of \revb{the anomaly amplitude} $q^\star$.
\revb{Besides the average Reynolds stress $\cal RS$ (eq.~\ref{eqdefRS})} we quantify the zonal wind geostrophy
with an rms $z$-length scale following \citet{Gastine2012} 
\begin{equation}
\ell_z= \frac{\left[\overline{u}_\phi\right]_{\mathrm{rms}}}{\left[\partial
\overline{u}_\phi / \partial
z \right]_{\mathrm{rms}}}\;\;.
\end{equation}
The impact of the imposed heat flux boundary pattern \revb{on the deep temperature structure}
is measured by the \revb{globally} averaged latitudinal gradient of the axisymmetric
temperature:
\begin{equation}
\Delta_\Theta = \left[\frac{\partial \overline{T}}{\partial \theta}
\right]_{\mathrm{rms}}.
\end{equation}
The kinetic energy of geostrophic and ageo\rev{s}trophic zonal
flows quantify the importance of the respective contributions:
\begin{eqnarray}
\geos{E}= \frac{1}{2V} \int \rev{\geos{\overline{u}_\phi}^2} dV \;\;, \\
\ageos{E} = \frac{1}{2V} \int \rev{\ageos{\overline{u}_\phi}^2} dV \;\;,
\end{eqnarray}
where $V$ is the volume of the spherical shell.

\begin{figure}
\includegraphics[width=0.96\columnwidth]{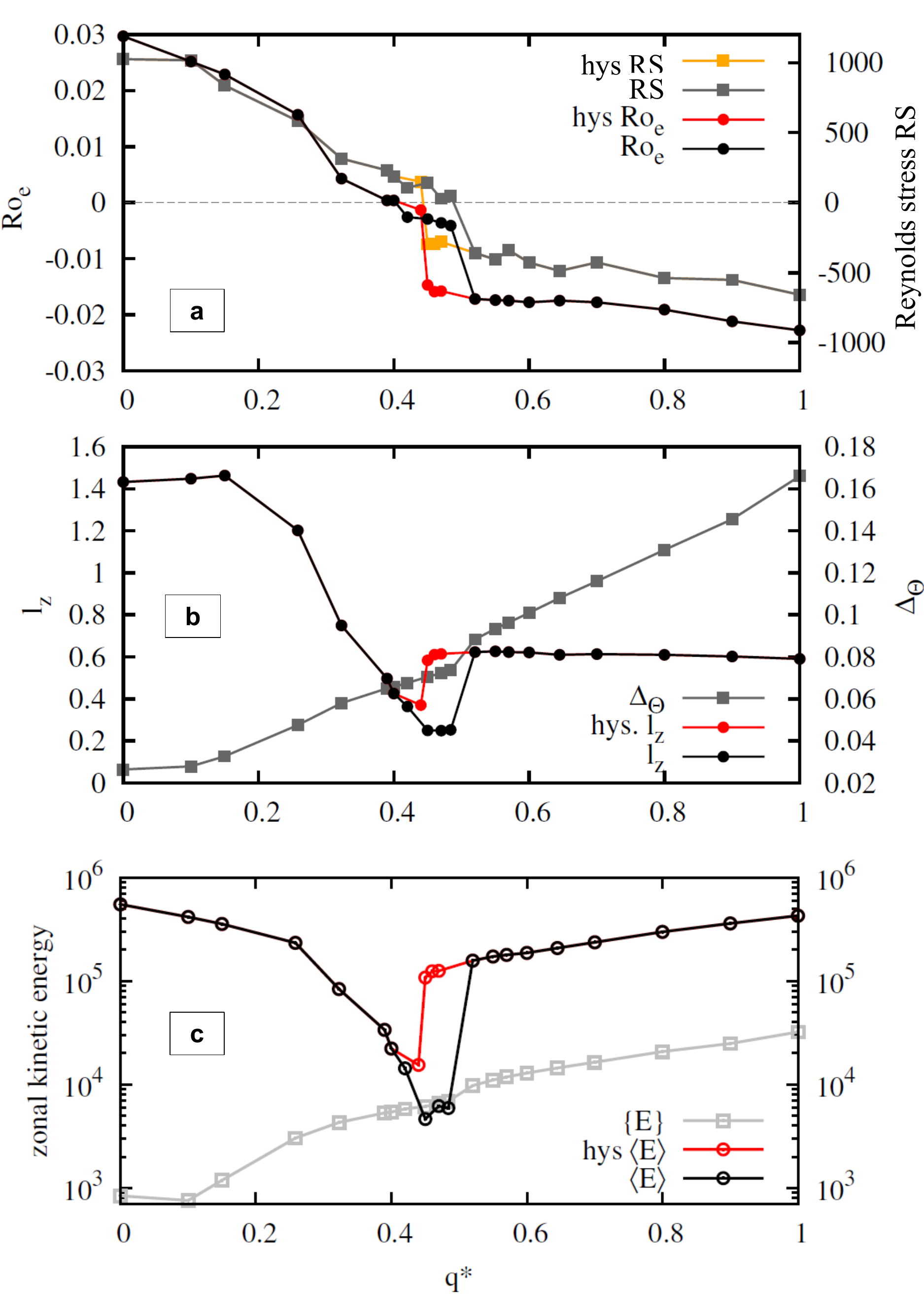}
\centering
\caption{Various flow properties as a function of the heat flux perturbation
amplitude $\qs$:  (a) zonal Rossby number at the equator $Ro_e$ and mean
Reynolds stress; (b) $\ell_z$ and mean latitudinal gradient of axisymmetric
temperature; (c) kinetic energy of zonal flows for geostrophic and ageostrophic
flow contributions. The orange and red profiles are calculated with decreasing
$\qs$ to visualise the hysteretic character.}
\label{mpvarq}
\end{figure}

Fig.~\ref{mpvarq} shows the variation of these different diagnostic
quantities with increasing $\qs$. The first panel of fig.~\ref{mpvarq}
demonstrates the clear correlation between the equatorial jet \revb{velocity} 
 and the Reynolds stress ${\cal RS}$.
This once more highlights that the Reynolds stress drives the
geostrophic zonal winds and that its inversion correlates with
 {the change in the zonal flow direction}. 

\revb{The length scale} $\ell_z$ \revb{and thus geostrophy} \revb{first} decreases with growing thermal wind but
recovers and saturates once the geostrophic wind direction has reversed.
Not surprisingly, $\Delta_\Theta$ increases roughly linearly with $q^\star$
as does the rms amplitude of ageostrophic mostly thermal wind related
zonal flows measured by $\ageos{E}$ in Fig.~\ref{mpvarq}c.
The geostrophic contribution, however, has a minimum around the
zonal flow reversal and clearly dominates at $\qs=0$ and $\qs=1$.
Once the convective columns have changed their tilt Reynolds stresses
efficiently drive the geostrophic zonal flows.

Interestingly, there is a hysteretic \revb{behaviour} between increasing and
decreasing $\qs$ that is also found in the average geostrophy of the zonal
flow (Fig.~\ref{mpvarq}b).
The hysteretic \revb{behaviour} can be found for all measures involving the
geostrophic zonal winds, which suggests that the zonal flow direction
indeed plays a role in determining the tilt.
Whichever mechanism ultimately changes the tilt has to overcome the
tilt direction promoted by the zonal wind shear.
A similar \revb{hysteretic behaviour} is also reported by \cite{Gastine2013b} who study the
zonal flow reversal happening at large Rayleigh numbers
for Rossby numbers around one.

\subsection{Rayleigh and Ekman number}
Rayleigh and Ekman numbers may have a crucial impact on the \revb{zonal} flow
\revb{direction and amplitude}. \revb{I}n the rotation-dominated convection regime,
 the Rayleigh number \revb{first increases with Rayleigh number simply because the convective flow amplitudes grow. For larger $Ra^\star$ Reynolds stresses decrease because the correlation between non-axisymmetric flow contributions is gradually lost \citep{Christensen2002b,Gastine2012}.}

\begin{figure}
\includegraphics[width=0.96\columnwidth]{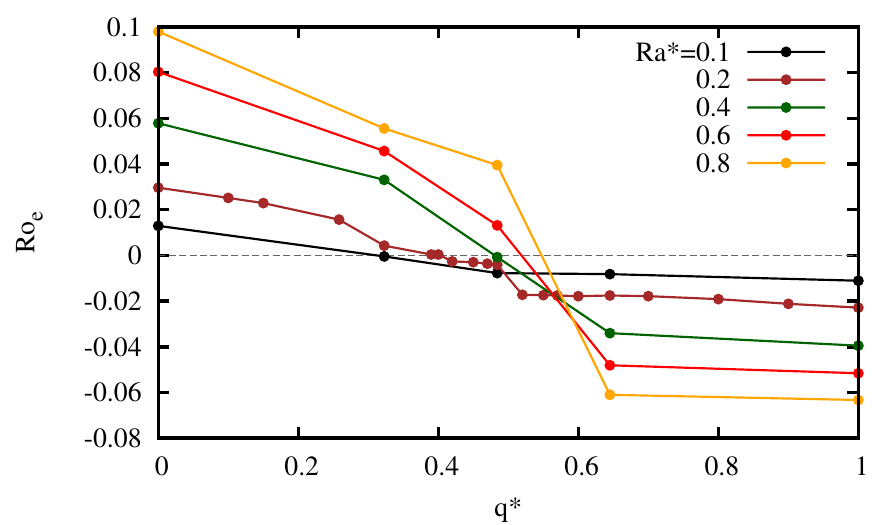}
\centering
\caption{Equatorial zonal flow $Ro_e$ at the outer boundary as a function of
the perturbation amplitude $q^\star$ for 5 different Rayleigh numbers.}
\label{ravarq}
\end{figure}

On the other hand, $Ra^\star$ is the
only parameter in the thermal wind balance (eq.~\ref{eqtw}) and therefore
\revb{controls the} ageostrophic zonal flows.
To explore the parameter dependence, we compute several
numerical models spanning the range from $Ra^\star=0.1$ to $Ra^\star=0.8$ with
a fixed Ekman number of $E=10^{-4}$. Figure~\ref{ravarq} shows \revb{how the equatorial zonal wind changes as $\qs$ is increased.}
\revb{In general, the amplitude of the equatorial jet for both directions scales with $Ra^\star$.}
\revb{The transition of the direction happens earlier for smaller $Ra^\star$.} 
\revb{When the zonal winds are weaker to start with, the boundary induced effect seems to have an easier task to change the direction.}

\revb{In the curve for $Ra^\star=0.2$, there is a pronounced jump in velocity when the jet is inverted as already discussed above. A similar jump is expected to happen for the other cases. }
Relying on Reynolds stresses,
the transition is a non-linear phenomenon and therefore particularly sensitive
to flow amplitudes and thus to Rayleigh numbers. More vigorous convection at higher
$Ra^\star$ leads to a more abrupt transition.

To investigate the impact of a smaller Ekman numbers $E$, we adjust the
Rayleigh number to provide roughly the same $Ro_e$ \revb{for the homogeneous reference case}  for all Ekman numbers. 
\revb{The ($E =10^{-4}$)-model served as a reference and we adjusted the $Ra^\star$-values accordingly.}

\begin{figure}
\includegraphics[width=0.96\columnwidth]{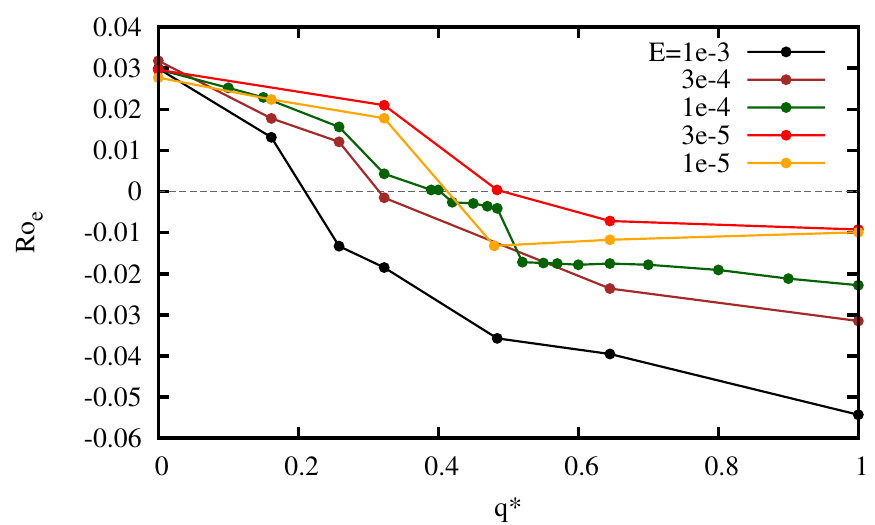}

\centering
\caption{Equatorial zonal flow $Ro_e$ at the outer boundary as a function of
the perturbation amplitude $q^\star$ for 5 different Ekman numbers.}
\label{ekvarq}
\end{figure}

Fig.~\ref{ekvarq} shows $Ro_e$ as a function of $q^\star$ for 5 different Ekman
numbers spanning the range from $E=10^{-3}$ to $E=10^{-5}$. Higher Ekman
numbers result in stronger reverted zonal flows at the equator, i.e.~higher $Ro_e$.
Smaller Ekman numbers promote more geostrophic flows and higher
perturbation amplitudes are required to overcome this constraint.
While $Ro_e(q^\star=0)$ is similar for all combinations of $Ra^\star$ and $E$
in Fig.~\ref{ekvarq}, $Ra^\star$ decreases from $0.6$ at $E=10^{-3}$ to
$0.08$ at $E=10^{-5}$. As the zonal flow amplitude in the \revb{inverted jet}
regime increases with $Ra^\star$ (Fig.~\ref{ravarq}), the low Ekman number
cases yield weaker equatorial jets. However, given the
crossover between cases for $E=3\cdot 10^{-5}$ (red) and $E=10^{-5}$
(orange) there might be another effect not understood so far.

\section{Discussion}
\label{sec:conclu}

We have shown that imposing an outer boundary heat flux pattern that
allows more heat to escape at higher latitudes while \revb{suppressing} convection
in the equatorial region can revert the geostrophic flow system.
Strongly retrograde equatorial jets like those observed on
Uranus or Neptune are the consequence.
The process is fairly general and happens at different Rayleigh and
Ekman numbers when the amplitude $\qs$ of the imposed heat flux variation
reaches about 50\% of the mean heat flux.

The imposed boundary condition is primarily responsible for
a deep reaching temperature anomaly that mirrors the imposed pattern:
\revb{t}he equatorial region remains hot while mid to high  latitude \revb{regions} are
cooled more efficiently by convection. Strong thermal winds are the
direct consequence of the respective latitudinal temperature gradient \revb{and as expected, these thermal wind speeds increase with $\qs$. However, the effects on the geostrophic flow, i.e.~ reversal and re-amplification for large enough positive $\qs$ as well as the boost for negative $\qs$ are indeed unforeseen.}

An analysis of the force balance has shown that the geostrophic zonal flows
are always driven by Reynolds stresses, i.e.~by a statistical
correlation of non-axisymmetric flow contributions. The axisymmetric
thermal wind has therefore no direct impact \revb{on} the
flow reversal. However, \revb{our analysis demonstrates} the reversal
correlates with a change in the tilt of convective columns.
This is perhaps not surprising, since this tilt is always
fundamental in establishing geostrophic Reynolds stresses,
but it offers the potential explanation: the imposed boundary
conditions could actually affect the tilt.

Unfortunately, we have not found a convincing connection so far and
may also face a chicken and egg problem here since the shear related
to zonal flows can by itself determine the tilt of convective columns.
Sure enough, Reynolds stress and columnar tilt are always consistent
in our simulations. \revb{The typical concepts for explaining the tilt and geostrophic zonal flow, such as the Rossby wave propagation \citep{Busse1982,Takehiro2008}, can not explain the numerical results.}

\revb{It seems tempting to apply our numerical results to the giant planets bearing the question what would be a reasonable outer boundary heat flux pattern. Here we consider the possible effect of intense solar irradiation. The heat flux from the deeper regions is likely reduced where the solar incident flux heats the outermost atmosphere more effectively. The details of this process depend on e.g., radiative transfer, albedo variations, convection in the weather layer and chemical processes. However, here we take a simpler approach and consider only the lateral distribution of the mean solar irradiation. Relevant here is the temporal mean of the solar irradiation over time scales required to alter zonal flows.}

\revb{Our numerical simulations, but also Jupiter's zonal flow structure suggest that the zonal winds generally change very slowly \citep{Vasavada05}. Since the zonal flows are predominantly geostrophic any change requires to accelerate a significant fraction of the total planetary mass and can therefore only be slow. The associated time scales $\tau_{ZF}$ can be estimated via the driving Reynolds stresses }
\begin{equation}
\frac{\partial \overline{v}_\phi}{\partial t} \approx \frac{\overline{v}_\phi}{\tau_{ZF}} \approx  \overline{ \vec{v}_c \cdot \nabla \vec{v}_c } \approx C \frac{v_c^2}{d_{ZF}}  \ , 
\end{equation}
\revb{yielding}
\begin{equation}
\tau_{zf} = \frac{\overline{v}_\phi \, d_{ZF} }{ C \, v_c^2} \ .
\label{eqtauzf}
\end{equation} 
\revb{Here $d_{ZF}$ is the width of the zonal flow jet and $C$ the mean correlation.  $\overline{v}_\phi$ and $v_c$ are the dimensional flow speeds of peak azimuthal and convective flow. } 

\revb{Assuming a perfect correlation $C=1$ provides an upper bound. Using for example Jupiter's zonal flow maximum of $\overline{v}_\phi=150\,\mathrm{m/s}$ observed by the Galileo entry probe, the width of the equatorial jet $d_{ZF} = 2.1 \cdot 10^6\, \mathrm{m}$ and an estimate for the convective flow speeds of $v_c=1 \,\mathrm{cm/s}$ \citep{Jones2014,Gastine2014} yields $\tau_{ZF} = 0.1\, \mathrm{Myr}$ or $8400$ orbital (sideric) revolutions. Equivalent estimates for Saturn, Neptune and Uranus can be found in tab.~\ref{tabtimes}.  }

%\revb{One possibility is the solar irradiation that would heat certain regions of the outermost atmosphere more than others. Where the atmosphere is hotter, the heat flux from the deeper atmospheric regions is likely reduced. For planets with smaller obliquity, like Jupiter, Earth, Saturn and Neptune ($3.12^\circ$, $23.44^\circ$, $26.73^\circ$, $29.56^\circ$, respectively) the equatorial regions receive more solar irradiation power than higher latitudes. For Uranus with a large axial tilt angle of $\approx 98^\circ$ this is less clear. }

\begin{figure*}
\includegraphics[width=0.6\textwidth]{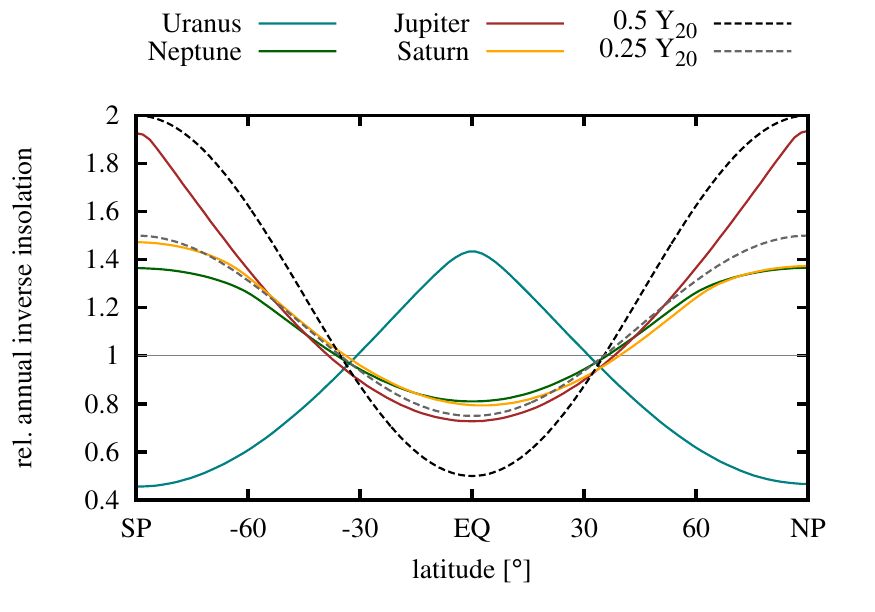}
\centering
\caption{Inverse normalised annual solar irradiation as a function of planetary latitude atop the atmospheres of the four giant planets and $Y_{20}$-patterns.}
\label{meansolir}
\end{figure*}
\revb{This implies that when calculating the impact of solar irradiation we have to consider the average over a sideric orbit. Fig.~\ref{meansolir} shows the respective inverse latitudinal insolation profiles for Jupiter, Saturn, Uranus and Neptune in comparison to $Y_{20}$ patterns as used in the study. Those are found by averaging daily irradiation pattern for each planet while taking orbital properties, such as obliquity and eccentricity, into account (accordingly to \citet{vanHemelrijck1982}). The profiles for Jupiter, Saturn and Neptune show an enhanced insolation (reduced internal heat flux) in the equatorial region. Due to the large obliquity, however, the pattern for Uranus is flipped since the high polar irradiation during summer dominates the annual mean \citep{vanHemelrijck1988}.} 

\revb{Observations suggest that the latitudinal total emission profiles of all four giant planets are rather flat \citep{Ingersoll1976,Pearl1990,Pearl1991,Soderlund2013}. Consequently the solar irradiation must be compensated either by a latitudinal variation of the internal heat flux \citep{Aurnou2008} or by equilibrating processes in the upper atmosphere. Adopting the former scenario allows to estimate the internal flux by simply inverting the irradiation profiles. Fig.~\ref{meansolir} shows that the $Y_{20}$-pattern with variable $\qs$ indeed provides a reasonable match for Jupiter, Saturn and Neptune. Uranus insolation however suggests a negative $\qs$ value.} 

%Although the effects of reflection and absorption in the outermost atmosphere are not covered in our models, it seems reasonable to assume a latitudinally flat profile of the total emission \citep{Ingersoll1976,Aurnou2008}. This is consequently modelled by a heat flux profile assumed at the upper edge of the optically thick atmosphere, which is inverse to the solar insolation (fig.~\ref{meansolir}). 

\revb{For Jupiter and Saturn the internal heat flux is roughly equal to the absorbed insolation \citep{Guillot2007}. Since the internal heat flux is also hugely superadiabatic, we can directly translate the inverse insolation pattern into a $\qs$-value. Fig.~\ref{meansolir} shows that for Jupiter a $\qs \approx 0.5$ is required to compensate the insolation pattern.  Our numerical simulations suggest that a mild heat flux variation of $\qs = 0.3$ generates a dimple of the equatorial prograde jet similar to Jupiter's main belt structure (see fig.~\ref{rozonvarq}). Equilibrating convection within the upper atmosphere may possibly reduce the effective $\qs$ for Jupiter's deeper convection.
For Saturn our approximation suggests $\qs \approx 0.25$, but since no dimple in the equatorial main jet has been observed the insolation flux maybe almost entirely compensated by upper atmosphere circulations. 

The insolation pattern also suggests $\qs\approx 0.25$ for Neptune. Since the internal heat flux is $50\%$ larger than the insolation flux, the effective $\qs$ reduces to ca.~0.15 \citep{Guillot2007}. This seems insufficient to create the inverse zonal flow patterns observed for Neptune. Our simulations at least require $\qs = 0.5$ for inverse zonal flow directions and possibly much larger values to also reach appropriate flow speeds.

However, for Uranus the peculiar insolation pattern exclude our proposed mechanism for explaining the jet directions. The fact that the emitted flux is almost equal to the absorbed flux means that the minuscule internal flux can not be the reason for the roughly homogeneous emission profile. At least for Uranus the convective processes within the upper atmosphere must eradicate any horizontal insolation gradient. 

Alternative attempts to explain the inverse differential rotation on the \revb{i}ce giants had to invoke the angular momentum mixing found at large Rayleigh numbers \citep{Aurnou2007, Gastine2013, Soderlund2013}. }

\revb{
\begin{table}[t]
\centering
\renewcommand{\arraystretch}{1.4} % Default value: 1
\begin{tabular}{c|cccc}
planet & $\overline{v}_\phi$ [m/s] & $d_{ZF}$ [$10^6$\,m] & $\tau_{ZF}$  [Myr] & $\tau_{sid}/\tau_{ZF}$ \\
\hline
Jupiter &  150  & 2.1 & 0.1 &  $8.4\cdot 10^3$ \\
Saturn &  450  & 6.2 & 0.87 & $2.9\cdot 10^4$ \\
Uranus &  200  & 3.0 & 0.19 &  $2.3\cdot 10^3$ \\
Neptune &  300  & 7.9 & 0.75 &  $8.1\cdot 10^3$ 
\end{tabular}
\caption{Characteristic time scales of zonal flow variations $\tau_{ZF}$ based on eq.~\ref{eqtauzf} in comparison to a sideric period $\tau_{sid}$. The jet width $d_{ZF}$ is estimated by the width of the equatorial jet, $v_\phi$ is the peak equatorial jet.} 
\label{tabtimes}
\end{table}}

\revb{Even though the proposed mechanism may not apply to the atmosphere of the planets in the solar system it is nevertheless an interesting hydrodynamic effect that requires further investigation. It is interesting that the common theories
of differential rotation in rapidly rotating spherical shell convection
fail to explain the effect of thermal inhomogeneities applied at the outer boundary.}

\section*{Acknowledgements}
\revb{The authors thank two anonymous referees for constructive suggestions significantly improving the manuscript.} 
WD is supported in part by the Science and Technology
Facilities Council (STFC), 'A Consolidated Grant in Astrophysical Fluids' (reference ST/K000853/1). JW and TG are supported by the Special Priority Program 1488 (PlanetMag, www.planetmag.de) of the German Science Foundation.

\bibliographystyle{elsarticle-harv}
\bibliography{mode20}

\end{document}